\documentclass[]{fairmeta}
\usepackage{amsmath}
\usepackage{graphicx}
\usepackage{booktabs}
\usepackage{hyperref}
\usepackage{xcolor}
\usepackage{algorithm}
\usepackage{algorithmic}
\usepackage{multirow}
\usepackage[table]{xcolor}

\usepackage{comment}
\newtoggle{comment}
\togglefalse{comment}
\newcommand{\jh}[1]{\iftoggle{comment}{\textcolor{orange}{[JH: #1]}}{}}

\tcbset{
    sharp corners,
    colback = white,
    before skip = 0.2cm,    
    after skip = 0.5cm      
}     

\newtcolorbox{boxA}{
    fontupper = \bf,
    boxrule = 1.5pt,
    colframe = black 
}

\title{Principled Synthetic Data Enables the First Scaling Laws for LLMs in Recommendation}
\author[1]{Benyu Zhang}
\author[1]{Qiang Zhang}
\author[]{Jianpeng Cheng}
\author[]{Hong-You Chen}
\author[]{Qifei Wang}
\author[]{Wei Sun}
\author[]{Shen Li}
\author[]{Jia Li}
\author[]{Jiahao Wu}
\author[]{Qunshu Zhang}
\author[]{Neeraj Bhatia}
\author[]{Xiangjun Fan}
\author[]{Hong Yan}

\affiliation[1]{Equal contributions}

\abstract{Large Language Models (LLMs) represent a promising frontier for recommender systems, yet their development has been impeded by the absence of predictable scaling laws, which are crucial for guiding research and optimizing resource allocation. We hypothesize that this may be attributed to the inherent noise, bias, and incompleteness of raw user interaction data in prior continual pre-training (CPT) efforts. This paper introduces a novel, layered framework for generating high-quality synthetic data that circumvents such issues by creating a curated, pedagogical curriculum for the LLM. We provide powerful, direct evidence for the utility of our curriculum by showing that standard sequential models trained on our principled synthetic data significantly outperform ($+130\%$ on recall@100 for SasRec) models trained on real data in downstream ranking tasks, demonstrating its superiority for learning generalizable user preference patterns. Building on this, we empirically demonstrate, for the first time, robust power-law scaling for an LLM that is continually pre-trained on our high-quality, recommendation-specific data. Our experiments reveal consistent and predictable perplexity reduction across multiple synthetic data modalities. These findings establish a foundational methodology for reliable scaling LLM capabilities in the recommendation domain, thereby shifting the research focus from mitigating data deficiencies to leveraging high-quality, structured information.}

\date{\today}
\correspondence{Benyu Zhang at \email{byzhang@meta.com}}

\metadata[Publication]{Accepted by ICML 2026}

\begin{document}

\maketitle

\section{Introduction}

Large Language Models (LLMs) represent a promising frontier for recommender systems, offering advanced sequence modeling capabilities and extensive world knowledge that can transcend the limitations of traditional embedding-table architectures~\cite{kangSelfAttentiveSequentialRecommendation2018, zhaiActionsSpeakLouder2024}. By representing users, items, and interactions within a unified linguistic space, LLMs can leverage textual descriptions to generate more sophisticated, content-aware, and explainable recommendations~\cite{rajputRecommenderSystemsGenerative2023, chenHLLMEnhancingSequential2024, hePLUMAdaptingPretrained2025}. The development of such systems at scale, however, requires the establishment of predictable scaling laws---indispensable instruments for navigating the substantial investments in data, computation, and engineering that modern LLM development demands~\cite{hoffmann2022training}. Yet, despite their established significance in NLP, \emph{no robust scaling laws
  have been established for the continual pre-training (CPT) of LLMs within the recommendation domain}~\cite{zhouOneRecTechnicalReport2025, hePLUMAdaptingPretrained2025}, compelling researchers to depend on costly trial-and-error methodologies.

  Prior work on scaling LLMs for recommendation has taken model-centric~\cite{yan2025unlockingscalinglawindustrial}, distillation-centric~\cite{Lai_2025}, or system-centric approaches, while still relying on raw user interaction logs as the primary data source. However, such logs are characterized by systemic flaws: noise, sparsity, and---most critically---pervasive biases including position bias, popularity bias, and exposure bias~\cite{chen2021biasdebiasrecommendersystem, NEURIPS2023_eb3c42dd}. Recent theoretical and empirical work has shown that scaling laws only manifest when data has sufficient quality and diversity~\cite{yang2024syntheticcontinuedpretraining, allenzhu2024physicslanguagemodels31}. Indeed, the PLUM framework~\cite{hePLUMAdaptingPretrained2025} encountered suboptimal scaling behavior where a 3B parameter model failed to consistently outperform its 900M counterpart---a clear symptom of training on information-poor, structurally biased data.

  The central insight of this work is that prior scaling failures were not primarily algorithmic but stemmed from a deficient data paradigm. Rather than designing models robust to data imperfections, we shift focus to constructing high-quality, ``pedagogical'' data engineered to teach the model the principles of recommendation in a structured manner. We introduce a novel, layered synthetic data framework that systematically decouples true user preference signals from system-induced artifacts. Layer~1 establishes foundational knowledge through item-text alignment and collaborative filtering data, while Layer~2 generates position-debiased user interaction histories via graph-based random walks that have no intrinsic concept of position or presentation order.

We provide powerful, direct evidence for the utility of this framework. First, standard sequential recommendation models (GRU4Rec \cite{hidasi2016sessionbasedrecommendationsrecurrentneural}, NARM \cite{li2017neuralattentivesessionbasedrecommendation}, STAMP \cite{gao2025stampscalabletaskmodelagnostic}, and SASRec \cite{kangSelfAttentiveSequentialRecommendation2018}) trained on our synthetic data achieve \textbf{significantly higher Recall@K} than models trained on real data across all cutoff points, demonstrating that our data captures more generalizable user preference patterns. Second, we empirically demonstrate, \textbf{for the first time}, robust power-law scaling for LLMs continually pre-trained on this data. Across model scales from 0.6B to 8B parameters trained on 163B tokens, we observe consistent scaling behavior characterized by the law $L(D) = L_\infty + A \cdot D^{-\alpha}$. Third, ablation studies reveal an asymmetric synergy between data layers: including collaborative filtering data alongside user interaction histories reduces asymptotic UIH loss by 31\% ($L_\infty = 0.66$ vs.\ $0.95$), while the reverse transfer does not hold---demonstrating that our layered curriculum provides
  complementary, non-redundant learning signals.

  Our quantitative analysis reveals a clear hierarchy of learning efficiency across data modalities. User Interaction History (UIH) data exhibits exceptional scaling exponents ($\alpha \approx 0.45$--$0.59$), indicating the model continues to benefit substantially from additional training tokens. Collaborative Filtering (CF) data shows strong scaling ($\alpha \approx 0.35$), followed by Item-Text alignment ($\alpha \approx 0.15$). In addition, our experiment has shown introducing CF data helped the learning of UIH data. These findings establish a foundational methodology for reliable, predictable LLM development in the recommendation domain, providing practitioners with the first quantitative roadmap for resource allocation and performance forecasting.

This paper makes the following principal contributions:
  \begin{itemize}
      \item \textbf{A layered synthetic data framework} that transforms noisy, biased user interaction logs into a high-quality curriculum. Layer~1 grounds semantic and collaborative knowledge through item-text alignment and association-rule data; Layer~2 generates position-debiased user interaction histories via graph-based random walks that eliminate positional and presentation-order artifacts.

      \item \textbf{Direct empirical validation of data utility}: We demonstrate that standard sequential models trained exclusively on our synthetic data outperform models trained on real data in downstream ranking tasks (Recall@10, @100, @1000), providing strong evidence that our curriculum captures more generalizable user preference patterns than raw interaction logs.

      \item \textbf{The first scaling laws for LLMs in recommendation}: Across model scales from 0.6B to 8B parameters trained on 163B tokens, we establish robust power-law scaling across seven evaluation domains. User interaction history exhibits the strongest scaling ($\alpha \approx 0.45$--$0.59$), followed by collaborative filtering ($\alpha \approx 0.35$) and item-text alignment ($\alpha \approx 0.15$), revealing a hierarchy of learning efficiency that provides a quantitative roadmap for data curation and resource allocation.
      \item \textbf{Discovery of asymmetric cross-domain transfer}: Ablation studies reveal that collaborative filtering data provides complementary signals that significantly improve user behavior modeling ($L_\infty$ reduced from $0.95$ to $0.66$, a 31\% improvement), while the reverse transfer does not hold. This validates the pedagogical design of our layered curriculum.
  \end{itemize}
\section{Related Work}
\subsection{Scaling Laws in Recommender Systems}
Scaling raw \cite{kaplan2020scalinglawsneurallanguage} characterize how a model's performance predictably (e.g., loss on evaluation set) improves as three key factors are increased: model size in the number of parameters ($N$); dataset size ($D$), or the number of training tokens; and the total compute ($C$ in FLOPS, which could be estimated as $C=6ND$ for transformer) used for training. The scaling law could be written as:
\begin{equation}
    L(N,D)=AN^{-\alpha}+BD^{-\beta}+E
\end{equation}
$L$ is the loss and $E$ is some constants. Chinchilla scaling raw \cite{hoffmann2022training} shows for a fixed compute budget, model size and dataset size must be scaled in tandem to achieve optimal performance.

However no such scaling law is yet established in recommendation models and the challenge of achieving predictable scaling for recommendation models has recently become a central focus of industrial and academic research. Several high-profile approaches have been proposed, each with a distinct philosophy. 

The Large User Model (LUM) paradigm from \cite{yan2025unlockingscalinglawindustrial} introduces a novel training task, "next-condition-item prediction," designed to better capture contextual user preferences. This approach is fundamentally model-centric, seeking to unlock scaling through a more sophisticated model architecture and training objective while still relying on raw user interaction logs as its primary data source. Another approach, the Stacked Unified Attention Network (SUAN) \cite{Lai_2025}, focuses on the task of Click-Through Rate (CTR) prediction. The authors first construct a highly accurate, scalable "teacher" model and then use knowledge distillation to transfer its capabilities to a smaller, lightweight "student" model that can be deployed online efficiently.

PLUM~\cite{hePLUMAdaptingPretrained2025} adapts pre-trained LLMs for industrial-scale generative recommendations through continual pre-training on synthetic user-item interaction data. While PLUM demonstrates the effectiveness of CPT for recommendation tasks, it does not systematically study scaling laws or investigate the impact of data mixture ratios on model performance. OneRec series ~\cite{zhouOneRecTechnicalReport2025, zhouOneRecV2TechnicalReport2025, liuOneRecThinkInTextReasoning2025} proposed a generative recommendation algorithm and studies their scaling behavior in the axis of model scales and test-time.

These prominent works are either model-centric, system-centric, or distillation-centric. Our work is orthogonal and complementary, proposing a fundamentally data-centric solution. It posits that even the most advanced architectures, such as LUM, would struggle to exhibit predictable power-law scaling if trained on the pathologically flawed data we identify as the root inhibitor. By first solving the data quality problem, our framework provides the necessary foundation upon which these and other architectural innovations can be built and scaled predictably.

\subsection{Principled Synthetic Data Generation for Recommendation}
Training LLM with synthetic data was used in understanding the training dynamics of LLM \cite{yang2024syntheticcontinuedpretraining,allenzhu2024physicslanguagemodels31,hronTrainingLanguageModels2024, wang2025datavalueagescaling, kang2025demystifyingsyntheticdatallm} and recently in industry LLM training \cite{yangQwen3TechnicalReport2025}. The use of synthetic data in recommender systems is an established field of research, though its application has traditionally focused on data augmentation \cite{qi2020searchbasedusermodelinglifelong} and privacy preservation \cite{adouani:hal-05113907} rather than enabling fundamental scaling properties. Early methods relied on statistical techniques to generate new data points that mirrored the distributions of the original dataset. Deep generative models such as Generative Adversarial Networks (GANs) \cite{bharadhwajRecGANRecurrentGenerative2018} and Variational Autoencoders (VAEs) \cite{adouani:hal-05113907} have been employed to create more sophisticated synthetic data, often with the goal of augmenting sparse datasets or creating privacy-preserving alternatives to real user data.

Our framework represents a conceptual departure from these traditional applications along three dimensions: (i)~\textit{GAN/VAE-based augmentation} (RecGAN~\cite{bharadhwajRecGANRecurrentGenerative2018}; Adouani et al.~\cite{adouani:hal-05113907}) replicates the source data distribution, inheriting its biases; our framework instead \textit{purifies} the signal by generating data from a debiased CF graph rather than mimicking flawed logs. (ii)~\textit{Sequence augmentation} methods (e.g., crop, mask, reorder~\cite{qi2020searchbasedusermodelinglifelong}) produce perturbed copies of existing sequences; our Layer~2 generates entirely new sequences via graph-based random walks, producing fundamentally new behavioral data rather than perturbed copies. (iii)~\textit{LLM-based generation} (e.g., using GPT to generate reviews or user profiles) operates at the sample level; our framework operates at the \textit{curriculum level}---designing structured training stages (semantics $\rightarrow$ collaborative structure $\rightarrow$ sequential behavior) that systematically build recommendation capabilities \cite{chen2025selfevolvingcurriculumllmreasoning, soviany2022curriculumlearningsurvey}. The key distinction is that prior methods aim to augment or expand existing data, while our framework constructs a pedagogical curriculum that decouples the learning of different recommendation concepts.
\subsection{Mitigating Bias in Recommendation Data}
The literature on debiasing recommender systems is vast, encompassing techniques such as inverse propensity scoring \cite{cardoso2022brsnisbiasreducedselfnormalized, agarwalEstimatingPositionBias2019, wangPositionBiasEstimation2018, Hu_2025}, causal inference \cite{pan2025counterfactualinferenceeliminatingsentiment, li2023stabledr}, and adversarial training \cite{yangAdversarialRegularizedDiffusion2025, feng2019learningfairrepresentationsadversarial, Hu_2023} to counteract the effects of systemic biases in training data. These approaches typically involve designing more complex model architectures or loss functions that can learn true user preferences despite being trained on confounded data.

Our work proposes a more direct and arguably more fundamental solution. In alignment with the broader "data-centric AI" movement \cite{zha2023datacentricartificialintelligencesurvey, angelakis2024datacentricapproachclassspecificbias}, we shift the engineering focus from the model to the data itself. Instead of building complex models to be robust to bias, we engineer position-debiased data. This allows for the use of powerful but standard LLM architectures without requiring bespoke modifications for debiasing, simplifying the modeling process and focusing resources on the creation of a high-quality, reliable source of truth for training.

\subsection{Graph-Based Methods for Recommendation}
GLTA~\cite{yang2025traininglargerecmodels} integrates collaborative filtering with LLMs via graph-language token alignment, treating user-item interactions as additional modalities to a graph neural network. While this is an interesting direction, whether it can scale to catalogs of hundreds of millions or billions of items remains unknown, and it may lack the reasoning capabilities that are an important advantage of LLMs for recommendation. S-Walk~\cite{choi2022swalkaccuratescalablesessionbased} explicitly models the inter-session relationship besides intra-session relations in random walks for session-based recommendation; its inter-session modeling could be supplementary to the Node2Vec method used in our Layer~2 design. Our framework differs fundamentally from both: rather than augmenting model architectures, we construct a pedagogical data curriculum that enables standard LLM training to exhibit predictable scaling.
\section{The Data Quality Bottleneck in Recommendation CPT}
\subsection{Fundamental Challenges in Adapting LLMs for Recommendation}
The adaptation of pre-trained LLMs for the recommendation domain presents several non-trivial challenges. A primary concern is real-time efficiency; the computational demands of large models can introduce significant latency, which is often unacceptable in production environments that require instantaneous recommendation generation. Another well-documented challenge is the cold-start problem, wherein the system struggles to provide accurate recommendations for new users or items with limited interaction histories.

However, arguably the most significant challenge is bridging the domain gap between the general-purpose knowledge encoded in a base LLM and the specific, often idiosyncratic, patterns of user behavior within a given recommendation domain. The model must learn to associate its understanding of language and concepts with the implicit signals embedded in user interaction sequences. The quality and nature of the data utilized for this domain adaptation are therefore of paramount importance.
\subsection{A Taxonomy of Systemic Biases in User Interaction Logs}
Raw user interaction logs, despite their volume, constitute a deeply flawed source of training data for a sophisticated pattern-matching system like an LLM. These logs are not a pure reflection of user preference but rather a distorted record heavily influenced by the very system that generated them \cite{chen2021biasdebiasrecommendersystem, NEURIPS2023_eb3c42dd, li2023fairerfairnessdecisionrationale, das2024unveilingmitigatingbiaslarge}. These data pathologies can be categorized as follows:
Table~\ref{tab:data_biases} categorizes these pathologies with concrete examples.

  \begin{table}[h]
  \centering
  \small
  \begin{tabular}{p{2cm}|p{9cm}|p{4cm}}
  \toprule
  \textbf{Bias Type} & \textbf{Description} & \textbf{Example} \\
  \midrule
  \textbf{Data Incompleteness \& Sparsity} &A typical user interacts with more than one platform. Furthermore, user profiles and item attributes are frequently incomplete, depriving the model of the rich contextual information required for nuanced reasoning. & User buys camera on Amazon, accessories on B\&H---neither platform sees full intent \\
  \midrule
  \textbf{Data Noise} & ser interactions are inherently noisy. A click may be accidental, a "like" may be reflexive, and a high numerical rating may be contradicted by a negative textual review. Such inconsistencies introduce conflicting signals that can degrade the quality of the model's learned representations. & 5-star rating with review: ``Arrived broken, returning it'' \\
  \midrule
  \textbf{Position Bias} & In ranked lists, users exhibit a strong tendency to interact with items presented at the top, irrespective of their absolute relevance \cite{bito2025evaluatingpositionbiaslarge,10.1145/3652864}. A model trained on such data will learn a simple, fallacious heuristic: "items in the first position are preferable." It learns to replicate the presentation bias of the preceding system, not the underlying preferences of the user. & CTR drops 50\%+ from position 1 to 5, even for equally relevant items \\
  \midrule
  \textbf{Popularity Bias} & Popular items are recommended more frequently, which leads to more interactions, which in turn reinforces their popularity \cite{braun2023metricspopularitybiasdynamic, 10.1145/3771279, 10.1145/3604915.3608825, Klimashevskaia_2024}. This feedback loop systematically under-represents and suppresses niche or long-tail content, leading to increasingly homogenized recommendations. & Top 1\% of items receive 80\% of interactions; long-tail items never surface \\
  \midrule
  \textbf{Exposure Bias} & Users can only interact with items they are shown . The interaction log is therefore a censored view of user preference, reflecting the past decisions and biases of the recommendation algorithm rather than the user's complete preference landscape \cite{mansoury2022exposureawarerecommendationusingcontextual, xu2025enhancinginterpretabilityeffectivenessrecommendation}. A model trained on this data cannot learn about items to which it was never exposed. & User loves jazz but system only recommends pop---logs show zero jazz preference \\
  \bottomrule
  \end{tabular}
  \caption{Taxonomy of systemic biases in user interaction logs. These artifacts are strong, consistent patterns that LLMs readily learn, mistaking them for genuine user intent.}
  \label{tab:data_biases}
  \end{table}

Continually pretraining an LLM on such data compels it to internalize these pathological patterns. The model's objective is to minimize its prediction error on the provided data, and the most efficient means to do so is to learn the dominant statistical regularities—which are the biases themselves.
\subsection{Analysis of Prior Art: The Sub-Scaling of PLUM}
The PLUM \cite{hePLUMAdaptingPretrained2025} framework serves as an illustrative case study of the consequences of training on raw user data. PLUM's methodology involves a CPT phase that utilizes domain-specific data, including user activity sequences, to adapt a pre-trained LLM for recommendation tasks. However, empirical results indicated that this approach faced significant scaling challenges. Specifically, the performance of the MoE-3B model was constrained by insufficient training data, ultimately preventing it from demonstrating clear superiority over the smaller MoE-900M model.

This outcome is a clear example of the "sub-scaling" phenomenon observed in the broader NLP literature. Recent analysis of scaling laws, such as the "Sub-Optimal Scaling Law" framework proposed in \cite{chen2025subscalinglawsroledata}, explicitly identifies high data density and redundancy as critical factors that cause performance gains to decelerate. When training data lacks diversity or contains significant redundancy \cite{tirumala2023d4improvingllmpretraining, he2024softdedupefficientdatareweighting}—characteristics typical of raw user interaction logs with their popularity-driven feedback loops—the marginal benefit of additional data diminishes rapidly. This aligns with the finding that models trained in high-density regimes deviate significantly from the anticipated power-law trajectory .

This leads to a critical insight regarding LLM training in this domain. When an LLM is continually pre-trained on biased user logs, it does not merely learn the bias; its powerful pattern-matching capabilities codify and amplify it. The model internalizes the system's flaws more effectively than simpler, traditional models. If this newly biased LLM is subsequently deployed, it will generate recommendations that are even more skewed, producing interaction data that is more contaminated than before. Using these new logs for a subsequent round of CPT would create a vicious cycle of bias amplification, where the system's quality degrades over time by continually learning and reinforcing its own errors. The only way to interrupt this degenerative loop is to fundamentally alter the nature of the training data itself.
\section{A Layered Framework for High-Fidelity Synthetic Data}\label{sec:data}
\subsection{Methodology: From Raw Logs to a Curated Curriculum}
To overcome the fundamental limitations of real-world interaction data, this work proposes a paradigm shift in data strategy. Rather than viewing data as a raw resource to be processed, our framework treats it as the foundation for a structured, pedagogical curriculum. The objective is not to mitigate biases in a flawed dataset but to construct an entirely new, high-fidelity dataset designed to teach the LLM the core concepts of the recommendation domain in a principled and progressive manner. This approach aligns with modern LLM development best practices, where models are trained on carefully curated and processed datasets rather than unfiltered web scrapes. For example \cite{yang2024syntheticcontinuedpretraining} found that continued pretraining with next-token prediction fails to teach the model to learn the knowledge its condensed representation, even with heavy repetition. Instead, generating a larger synthetic dataset grounded by the source documents can enable a continually pretrained model to learn the knowledge. However, for synthetic CPT to scale, the synthetic data must be sufficiently diverse. \cite{allenzhu2024physicslanguagemodels31} also showed diversity of the data (e.g., rewriting in different format) is the critical to the success of (continual) pre-training and merely repeating the data won't work. This framework implements a formal curriculum, organized into two primary layers that progress from foundational concepts (semantics and collaborative logic) to more complex, integrated patterns (full sequential behavior), with each layer designed to impart a specific type of knowledge.

\subsection{Layer 1: Grounding Semantic and Collaborative Knowledge}
The first layer of the curriculum furnishes the LLM with the foundational "vocabulary" and "grammar" of the recommendation domain. It teaches the model what items are and how they relate to one another, both through their intrinsic properties and through the collective behavior of users.
\subsubsection{Item-Text Alignment Data}
\textbf{Purpose}: The primary objective of this data type is to establish a robust semantic link between the abstract identifiers used to represent items (e.g., semantic tokens) and their real-world meaning as conveyed through textual descriptions. This step is critical for transcending the pure memorization of ID-based co-occurrence patterns that characterizes traditional collaborative filtering models. By grounding the item representations in language, the model can reason about content-based relationships, enabling superior generalization to new, long-tail, or cold-start items.

\textbf{Methodology}: The generation process involves creating direct mappings between an item's tokenized representation and its descriptive text. These pairs are formatted as natural language sequences. For example:
\begin{boxA}
This item <RECTOKEN> REC6594 REC5411 REC1547 REC941 REC7587 REC7639 REC3383 REC6576
</RECTOKEN> is described as [redacted] lotus yoga seed bead bracelet, by Handmade in
Women › Jewelry › Necklaces.
\end{boxA}
Here `<RECTOKEN>` and `</RECTOKEN>` are added special tokens to deliminate the semantic tokens; `REC6594` are added tokens for semantic tokens vocab. The textual descriptions can be sourced from product catalogs, content creator metadata, or synthesized by other LLMs to ensure richness and consistency. This data provides the model with the fundamental building blocks of content understanding.
\subsubsection{Collaborative Filtering (CF) Data}
\textbf{Purpose}: This data type aims to explicitly teach the model the statistical logic of collaborative filtering by translating user behavior patterns into a linguistic format that the LLM can process. Instead of requiring the model to implicitly infer these patterns from noisy sequences, this data makes the relationships explicit, providing a clear and direct learning signal.

\textbf{Methodology}: The process begins by mining association rules from the raw user interaction histories and existing item-to-item (I2I) data, converting these statistical relationships into templated natural language statements. For example, a basic association might be formatted as:

\begin{boxA}
When a user interacts with item <RECTOKEN> REC3078 REC3311 REC7479 REC2862 REC5552 REC4015
REC6914 REC4637 </RECTOKEN>, there is a 4.9\% probability they will also interact with item
<RECTOKEN> REC3078 REC3311 REC7479 REC4015 REC5211 REC2862 REC1723 REC6914 </RECTOKEN>
(confidence: 0.049, lift: 652.45)
\end{boxA}

\subsection{Layer 2: Simulating Position-Debiased User Behavior}
The second layer of the curriculum builds upon the foundational knowledge established in Layer 1. It utilizes the learned relationships to generate complete interaction sequences that are realistic yet free from the contaminating artifacts present in raw data.
\subsubsection{Synthetic User Interaction Histories (UIH)}
\textbf{Purpose}: The objective of Synthetic UIH is to generate clean, diverse, and privacy-preserving user interaction sequences. This data directly addresses the core problems of missing positives and position bias found in real logs. As the sequences are generated algorithmically, they are not tied to any individual user's real history, thus preserving privacy.

\textbf{Methodology}: The generation process employs a novel CF-to-LLM training pipeline. First, a graph is constructed where items are represented as nodes, and the weighted edges between them are determined by the strength of the collaborative filtering relationships mined in Layer 1. Realistic user journeys are then simulated using 2nd-order biased random walks, following the Node2Vec \cite{grover2016node2vecscalablefeaturelearning} algorithm. Unlike a 1st-order walk where the next step depends only on the current node, a 2nd-order walk's transition probability is conditioned on both the current node $v$ and the previous node $t$. The unnormalized transition probability $\pi_{v,x}$ to a next node $x$ is given by $\pi_{v,x} = \alpha_{p,q}(t, x) \cdot w_{v,x}$, where $w_{v,x}$ is the static edge weight and $\alpha_{p,q}(t, x)$ is the Node2Vec search bias. Return parameter $p$ controls the likelihood of returns to previous node $t$ and in-out parameter $q$ controls exploration ($q>1$) vs exploitation ($q<1$).

This layered approach functions as a powerful data purification process. It systematically decouples the true user preference signal from the system-induced artifacts. Layer 1, particularly the CF Data component, acts as an initial filter by extracting the core statistical signal of item co-occurrence from the noisy and biased raw logs. This process inherently averages over numerous user sessions, rendering it less susceptible to individual instances of position or presentation bias. Layer 2 then utilizes this purified signal as the sole basis for generating entirely new behavioral data. The random walk generation process has no intrinsic concept of "rank," "position," or "presentation order," ensuring that the resulting Synthetic UIH data is free from positional and presentation-order artifacts. We note that while this process eliminates position and temporal-ordering biases by construction, popularity bias encoded in graph edge weights may persist; we provide a quantitative bias audit in Appendix~\ref{app:bias_audit}. This two-stage filtering mechanism is the core mechanic that enables the creation of a clean, high-quality training curriculum, preserving valuable collaborative information while discarding contaminating biases. For example:
\begin{boxA}
A user interacted with the following sequence of items:
<RECTOKEN> REC870 REC5932 REC6271 REC1852 REC1624 REC409 REC6034 REC3608 </RECTOKEN>,
<RECTOKEN> REC870 REC5524 REC180 REC4637 REC5552 REC6862 REC6033 REC4948 </RECTOKEN>,
<RECTOKEN> REC7402 REC1581 REC5202 REC5289 REC2325 REC5067 REC6960 REC73 </RECTOKEN>,
<RECTOKEN> REC1815 REC4896 REC2334 REC7479 REC4502 REC4861 REC1295 REC6855 </RECTOKEN>,
<RECTOKEN> REC7479 REC1815 REC2334 REC3927 REC7667 REC2958 REC6513 REC4896 </RECTOKEN>,
<RECTOKEN> REC1815 REC4224 REC5068 REC2334 REC4861 REC6855 REC7766 REC2410 </RECTOKEN>,
<RECTOKEN> REC1815 REC2633 REC2433 REC2372 REC650 REC4064 REC2334 REC1295 </RECTOKEN>,
<RECTOKEN> REC1815 REC5412 REC6298 REC3831 REC6513 REC138 REC5680 REC4210 </RECTOKEN>
\end{boxA}
\section{Empirical Validation of Synthetic Data Quality}
The central thesis of this work is that high-quality synthetic data is the key to unlocking scaling laws. To empirically substantiate this claim, we evaluate the generated data across three dimensions: fidelity, utility, and privacy.
\subsection{Statistical Fidelity Analysis}
Fidelity measures how closely the synthetic data mirrors the statistical properties of the original data. We compared the distributions of key characteristics, such as item popularity and sequence length, between the generated Synthetic UIH and the original Merrec user logs.
\subsection{Empirical Validation of Data Utility for Downstream Ranking}
To assess the practical utility of our synthetic data for the core recommendation task of ranking, we conduct a "Train on Synthetic, Test on Real" (TSTR) evaluation and compare it against a "Train on Real, Test on Real" (TRTR) baseline. To ensure a fair comparison, this evaluation uses a constrained methodology: the test set, drawn from real user interactions, is filtered to include only items present in the vocabulary of both the synthetic and real training sets.

We trained four standard sequential recommendation models (GRU4Rec, NARM, STAMP, and SASRec) in both the TSTR (Syn→Real) and TRTR (Real→Real) settings. The results, visualized in Figure~\ref{fig:tstr_vs_trtr_recall} below, reveal a striking and powerful finding. In all cases, models trained exclusively on our synthetic data (Blue lines) achieve significantly better ranking performance (Recall@K) than models trained on real data (Red lines) across all cutoff points (@10, @100, @1K).
\begin{figure}[h]
    \centering
    \includegraphics[width=0.9\textwidth]{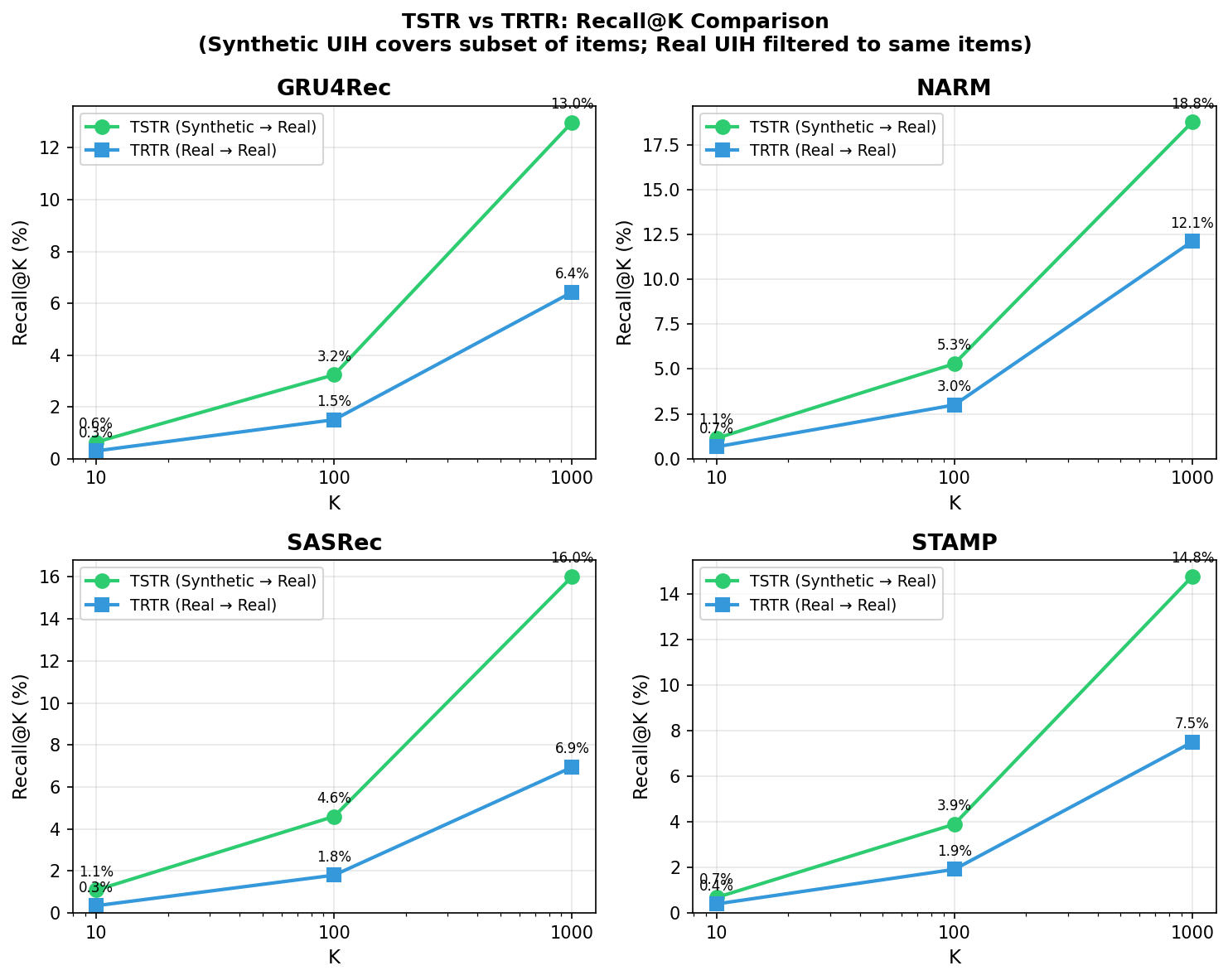}
    \caption{TSTR vs TRTR: Recall@K Comparison Across Models. TSTR (Train on Synthetic, Test on Real) consistently outperforms TRTR (Train on Real, Test on Real) across all models (GRU4Rec, NARM, SASRec, STAMP) and K-values. Note: Real UIHs are filtered to the same set of items as Synthetic UIHs.}
    \label{fig:tstr_vs_trtr_recall}
\end{figure}

These results provide strong, direct evidence that our synthetic data is not only a high-fidelity substitute but is in fact a superior training resource for teaching the general principles of user preference required for high-quality recommendation. By learning from the purified collaborative filtering signal rather than the noisy specifics of raw interaction logs, the models trained on synthetic data acquire more generalizable co-occurrence patterns. This leads to superior performance on ranking tasks, which require an understanding of the broad set of plausible next items rather than just the single most likely one.
\subsection{Privacy}
By design, our data generation process offers inherent privacy benefits. The Synthetic UIH sequences are generated from an aggregated graph model of item relationships, not from individual user histories. This ensures that the training data is decoupled from any single user's activity, mitigating privacy risks associated with using raw interaction logs.

\section{Principled Scaling Laws for Recommendation}
\subsection{Experimental Configuration}
Having established the superior utility of our synthetic data for downstream tasks, we now turn to our central hypothesis: that this high-quality data enables predictable scaling laws for LLMs. To empirically validate this, a series of continual pre-training experiments were conducted with the following settings:
\begin{itemize}
    \item \textbf{Base Model}: We use Qwen3 models scaling from 0.6B to 8B parameters (0.6B, 1.7B, 4B and 8B), which is one of the best open-weight LLM model available now;
    \item \textbf{Dataset}: Table\ref{tab:dataset}  describes the statistics of our training dataset.
\begin{itemize}
    \item \textbf{General}: We use three subsets (cosmopedia-v2, fineweb-edu-dedup and python-edu) from SmolLM-Corpus \cite{smollm2024} as our general domain data to avoid catastrophic forgetting in continual pretraining.
    \item \textbf{Recommendation}: The source for recommendation data was the merrec dataset \cite{merrec2024}, a publicly available C2C (consumer-to-consumer) e-commerce dataset from HuggingFace with 1.2B interactions from 65.7M items and 2.6M users. We use sparse auto-encoder generated semantic ID to represent the item and experiments around semantic ID is included in Appendix~\ref{app:token}
\end{itemize}
    \item \textbf{CPT Hyperparameters}: Continual pre-training was performed on B200 clusters using a global batch size = 512 sequences, context windows = 512 tokens, learning rate (peak at 1e-4) with a linear warmup (100 steps) and cosine decay schedule. All models would be trained with 163B tokens, which is over 20x (chinchilla scaling multiplier \cite{hoffmann2022training}) for 8B models.
    \item \textbf{Evaluation Metric}: Model performance was measured using perplexity on a held-out test set across all domains (general domain, item-text, CF both seen, CF one unseen, CF both unseen, UIH OOD, and UIH full graph). The statistics are presented in Table~\ref{tab:test_dataset}.
\begin{itemize}
    \item \textbf{General domain}: We randomly sample 200K entries (102.4M tokens) for evaluation purpose.
    \item \textbf{Item-text}: item-text pairs for items not in training set. We then sample 200K entries (102.4M tokens) for evaluation purpose.
    \item \textbf{CF both seen}: collaborative filtering edge between two items with items presented in training data but edges are not included in training data. We then sample 200K entries (102.4M tokens) for evaluation purpose.
    \item \textbf{CF one unseen}: CF edge between two items with one item presented in training data and the other items not. Obvious those edges won't present in training data. We then sample 200K entries (102.4M tokens) for evaluation purpose.
    \item \textbf{CF both unseen}: CF edge between two items with neither items nor edges presented in training data. There are about 75k entries (38M tokens) in total.
    \item \textbf{UIH OOD}: UIH sampled only based on edges from CF test set. There are about 48.3K entries (24.7M tokens). For the parameters used to generate this UIH, please refer to Appendix~\ref{sec:uih_parameter}
    \item \textbf{UIH full graph}: UIH sampled from both CF train and CF test set. There are about 51.7K entries (26.5M tokens).
\end{itemize}

\end{itemize}

\begin{table}[]
\caption{Statistics of the dataset used in this experiment. Our dataset consists of both general domain data \cite{smollm2024} and recommendation data from Merrec \cite{merrec2024} We mixes those two domain with a 50\% and 50\% ratio to avoid catastrophic forgetting \cite{ibrahim2024simplescalablestrategiescontinually} Within recommendation data, it is consisted of three types synthetic data from two layers as described in Section \ref{sec:data}. All models would be trained with 163B tokens and last row estimated the number of repeats each domain would be exposed after the training.}
\label{tab:dataset}
\centering
\begin{tabular}{l|llll}
\toprule
Domain & General & Item Text & CF     & UIH    \\
\midrule
Token  & 128B    & 2.0B    & 7.6B & 2.8B \\
Ratio  & 50\%    & 9\%       & 30\%   & 11\%  \\
Repeats & $<1$ & 6.2 & 7.3 & 6.1 \\
\bottomrule
\end{tabular}
\end{table}

\begin{table}[]
\caption{Statistics of the holdout test dataset used in this experiment, which is consisted of four domains: general, item-text, CF and UIH OOD.}
\label{tab:test_dataset}
\centering
\begin{tabular}{l|llll}
\toprule
Domain & General & Item Text & CF     & UIH    \\
\midrule
Token  & 102.4M    & 102.4M    & 243M & 51M \\
\bottomrule
\end{tabular}
\end{table}
\subsection{Power-Law Scaling Across Data Modalities}
\begin{figure*}[!t]
    \centering
  \begin{subfigure}{0.48\textwidth}
    \includegraphics[trim={0 680 859 0}, clip, width=1\textwidth]{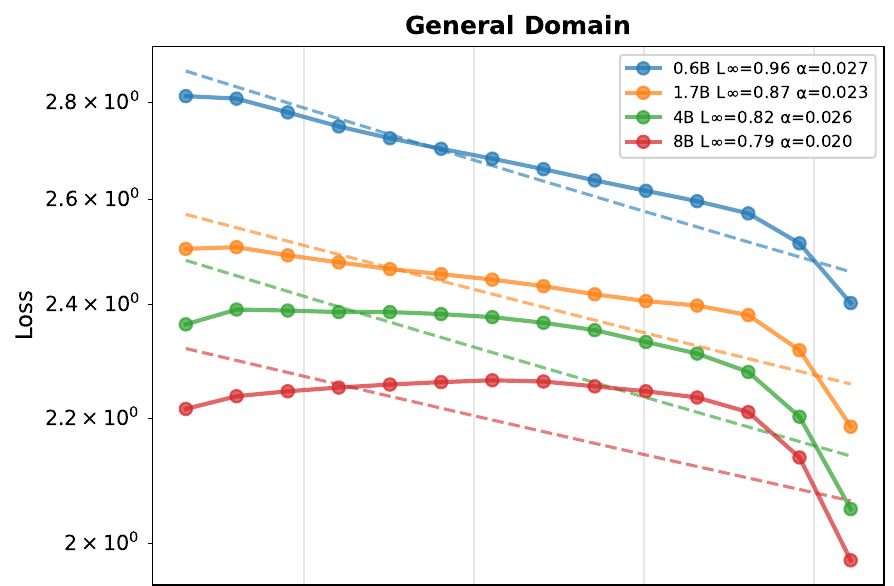}
  \end{subfigure}%
  \hfill
  \begin{subfigure}{0.48\textwidth}
    \includegraphics[trim={859 680 0 0}, clip, width=1\textwidth]{figure/socae_across_scales_30M.pdf}
  \end{subfigure}%
  \vfill
  \begin{subfigure}{0.37\textwidth}
    \includegraphics[trim={0 360 859 320}, clip, width=1\textwidth]{figure/socae_across_scales_30M.pdf}
  \end{subfigure}%
  \hfill
  \begin{subfigure}{0.31\textwidth}
    \includegraphics[trim={501 360 429 320}, clip, width=1\textwidth]{figure/socae_across_scales_30M.pdf}
  \end{subfigure}%
  \hfill
  \begin{subfigure}{0.31\textwidth}
    \includegraphics[trim={930 360 0 320}, clip, width=1\textwidth]{figure/socae_across_scales_30M.pdf}
  \end{subfigure}%
  \vfill
  \begin{subfigure}{0.54\textwidth}
    \includegraphics[trim={0 0 859 641}, clip, width=1\textwidth]{figure/socae_across_scales_30M.pdf}
  \end{subfigure}%
  \hfill
  \begin{subfigure}{0.45\textwidth}
    \includegraphics[trim={501 0 429 641}, clip, width=1\textwidth]{figure/socae_across_scales_30M.pdf}
  \end{subfigure}%
    \caption{Scaling laws for different domains across different model scales. UIH exhibits strong scaling ($\alpha_{\text{UIH}} = 0.45$--$0.59$), indicating continued improvement with additional training tokens. General domain shows near-saturation ($\alpha < 0.1$) as expected for pretrained models. The dashed lines are the fitted scaling law curves and parameters are provided as legend of the curve and Table~\ref{tab:model_scale}.}
    \label{fig:model_scale}
\vspace{-10pt}
\end{figure*}

The primary results are presented in Figure~\ref{fig:model_scale}, which depicts perplexity as a function of training tokens for seven data modalities. Both axes use logarithmic scales. Across all conditions—General Domain, Item-Text, CF Both Seen, CF One Unseen, CF Both Unseen, UIH OOD,
  and UIH Full Graph—the empirical data points follow remarkably consistent power-law trends.

  \subsubsection{Per-Model Data Scaling Analysis}

  We first fit per-model scaling laws of the form $\ell(D) = L_\infty + A D^{-\alpha}$, where $L_\infty$ is the ideal loss and $\alpha$ is the scaling exponent measuring the rate of perplexity reduction per unit of training data. The results are summarized in Table~\ref{tab:model_scale}.

  The scaling exponent $\alpha$ reveals a clear hierarchy of \emph{data efficiency} across modalities. UIH exhibits the strongest scaling ($\alpha \approx 0.45$--$0.59$), followed by CF ($\alpha \approx 0.28$--$0.36$), Item-Text ($\alpha \approx 0.13$--$0.21$), and General Domain ($\alpha
  \approx 0.02$--$0.03$). The near-saturation of General Domain is expected: the pretrained Qwen3 checkpoint already encodes substantial general-domain knowledge, leaving little room for improvement through continued pretraining on similar data.

  Two model-size-dependent patterns emerge. First, larger models exhibit higher $\alpha$ for CF and UIH, indicating that increased model capacity improves the ability to absorb new recommendation knowledge from data. Second, despite this trend, $L_\infty$ remains nearly constant across model
  sizes for CF and UIH, suggesting that even 0.6B parameters are sufficient to represent the underlying patterns in these domains at convergence—the bottleneck is data coverage, not model capacity.

  \subsubsection{Joint Scaling Law over Model Size and Data}

  To disentangle the contributions of model size ($N$) and training data ($D$), we fit joint scaling laws of the form $\ell = E + A \cdot N^{-\alpha} + B \cdot D^{-\beta}$, where $E$ is the irreducible loss, $\alpha$ governs model-size scaling, and $\beta$ governs data scaling:
  \begin{align}
      \ell_{\text{general}} &= 0.79 + 16500\, N^{-0.511} + 3.85\, D^{-0.048} \\
      \ell_{\text{item-text}} &= 0.473 + 5720\, N^{-0.511} + 4.04\, D^{-0.070} \\
      \ell_{\text{cf}} &= 0.193 + 21\, N^{-0.277} + 6.79\, D^{-0.148} \\
      \ell_{\text{UIH}} &= 0.514 + 1.89\, N^{-0.138} + 63.9\, D^{-0.272}
  \end{align}

  The joint fit yields several findings that complement and deepen the per-model analysis.

  \paragraph{Text and recommendation domains exhibit opposite scaling bottlenecks.}
  There is a striking inverse relationship between $\alpha$ (model-size exponent) and $\beta$ (data exponent) across domains:
  \begin{itemize}
      \item \textbf{General/Item-Text}: high $\alpha$ (0.511), low $\beta$ (0.048--0.070). These domains are \emph{model-size dominant}—performance improves primarily by scaling up model capacity.
      \item \textbf{UIH}: low $\alpha$ (0.138), high $\beta$ (0.272). This domain is \emph{data dominant}—performance improves primarily by increasing data coverage.
      \item \textbf{CF}: intermediate on both axes ($\alpha = 0.277$, $\beta = 0.148$), representing a balanced regime.
  \end{itemize}
  This pattern is consistent with the per-model observation that $L_\infty$ for CF and UIH varies little across model sizes: the joint fit explains this through the small $A$ coefficients for these domains (CF: 21, UIH: 1.89), compared to text domains (General: 16{,}500, Item-Text: 5{,}720).
  The model-size-dependent loss term $A N^{-\alpha}$ contributes minimally for recommendation domains, confirming that their bottleneck lies in data rather than model capacity.

  For text domains, the high $\alpha$ but low $\beta$ reflects two complementary factors: (1) the pretrained checkpoint already contains extensive general-domain knowledge, making additional text data marginally redundant; and (2) language modeling involves complex compositional patterns that
   inherently benefit from increased model capacity.

  \paragraph{Compute-optimal allocation differs drastically by domain.}
  Given a fixed compute budget $C = 6ND$, minimizing $\ell$ subject to this constraint yields the optimality condition $\alpha A \cdot N^{-\alpha} = \beta B \cdot D^{-\beta}$, requiring that the marginal loss reduction per FLOP be equal for model size and data. The coefficients $\alpha A$ and
   $\beta B$ (Table~\ref{tab:scaling_trade}) reveal sharply different optimal strategies: General and Item-Text domains should allocate compute overwhelmingly toward model size, while UIH should prioritize data collection. Concretely, halving the data-dependent loss term for General requires
  $2^{1/0.048} \approx 1.8 \times 10^6$ times more data, whereas for UIH it requires only $2^{1/0.272} \approx 13$ times more data.

  \begin{table}[t]
  \caption{Compute-optimal trade-off coefficients. Minimizing loss $\ell = E + AN^{-\alpha} + BD^{-\beta}$ subject to compute budget $C = 6ND$ yields the optimality condition $\alpha A \cdot N^{-\alpha} = \beta B \cdot D^{-\beta}$. The ratio of $\alpha A$ to $\beta B$ determines whether
  compute is best allocated to model size or data.}\label{tab:scaling_trade}
  \centering
  \begin{tabular}{l|l|l|l}
  \toprule
  Domain    & $\alpha A$ & $\beta B$ & Implication \\
  \midrule
  General   & 8{,}432  & 0.185  & Overwhelmingly model-size dominant \\
  Item-Text & 2{,}923  & 0.283  & Strongly model-size dominant \\
  CF        & 5.82     & 1.005  & Moderately model-size dominant \\
  UIH       & 0.261    & 17.38  & Strongly data dominant \\
  \bottomrule
  \end{tabular}
  \end{table}

  \paragraph{Irreducible loss reflects intrinsic predictability.}
  The irreducible loss $E$ captures the entropy floor of each domain:
  \begin{itemize}
      \item \textbf{CF} ($E = 0.193$): Most predictable, consistent with the low-rank structure of user-item co-occurrence matrices.
      \item \textbf{Item-Text} ($E = 0.473$): Moderately predictable, reflecting the constrained vocabulary and repetitive structure of item descriptions.
      \item \textbf{UIH} ($E = 0.514$): User behavior sequences exhibit moderate stochasticity—users are somewhat predictable but not deterministic.
      \item \textbf{General} ($E = 0.79$): Natural language has the highest inherent entropy, as expected.
  \end{itemize}

  \paragraph{CF and UIH capture fundamentally different recommendation signals.}
  Despite both encoding user behavior, CF and UIH exhibit distinct scaling profiles. CF scales better with model size ($\alpha = 0.277$ vs.\ $0.138$) but less with data ($\beta = 0.148$ vs.\ $0.272$). This distinction reflects their structural differences: CF encodes global user-item
  co-occurrence patterns, which benefit from model capacity to learn richer latent factors, while UIH encodes sequential, individual-level interaction trajectories, where the primary bottleneck is \emph{observational coverage}—seeing enough diverse user histories—rather than model
  expressiveness.

  \paragraph{Consistency between the two analyses.}
  The per-model and joint scaling analyses yield consistent conclusions through complementary lenses. The per-model fits establish the data-efficiency hierarchy (UIH $>$ CF $>$ Item-Text $>$ General), while the joint fit reveals the complementary model-size dimension and explains \emph{why}
  the hierarchy exists: domains where data scaling is efficient (high $\beta$) tend to have weak model-size scaling (low $\alpha$), and vice versa. The per-model observation that $L_\infty$ is stable across model sizes for CF and UIH is explained by the small $A$ coefficients in the joint
  fit. However, the per-model analysis also reveals that $\alpha_{\text{data}}$ increases with model size for CF and UIH, suggesting a model-data interaction that the additive joint formulation does not fully capture—an avenue for future investigation.

\begin{table}[h]
\caption{Scaling laws for different domains across different model scales. UIH exhibits strong scaling ($ \alpha_{\text{UIH}} = 0.453--0.689$), indicating continued improvement with additional training tokens. General domain shows near-saturation ($\alpha \approx0.025$) as expected for pretrained models. $L_\infty$ should be the lower the better and $\alpha$ should be the higher the better.}
\label{tab:model_scale}
\centering
\small
\begin{tabular}{lllllllll}
\toprule
      & \multicolumn{2}{c}{General}                      & \multicolumn{2}{c}{Item}   & \multicolumn{2}{c}{CF}     & \multicolumn{2}{c}{UIH}    \\
Model & $L_\infty\ \downarrow$ & $\alpha\ \uparrow$ & $L_\infty\ \downarrow$ & $\alpha\ \uparrow$ & $L_\infty\ \downarrow$ & $\alpha\ \uparrow$ & $L_\infty\ \downarrow$ & $\alpha\ \uparrow$ \\
\midrule
0.6B  & 0.96 &  \textbf{0.027} &   1.07 &  \textbf{0.169} &   0.33 &  0.282 & 0.64 &  0.453 \\
1.7B  &  0.87 &  0.023 &  0.92 &  0.141 & 0.34 & 0.315 & 0.65 &  0.514 \\
4B    &  0.82 & 0.026 &   \textbf{0.86} &  0.126 & 0.35 &  0.357 &  0.66 &  0.564 \\
8B    & \textbf{0.79} & 0.020 &  0.93 & 0.214 &  0.36 &   \textbf{0.358} &  0.66 & \textbf{0.588} \\
\bottomrule
\end{tabular}
\end{table}

\section{Scaling Analysis and Future Directions}
\subsection{Ablation Study: The Synergistic Effect of Layered Data}
A core tenet of our methodology is that the different layers of synthetic data are complementary, each teaching a distinct and necessary facet of the recommendation domain. To validate this hypothesis and quantify the contribution of each layer, we conduct an ablation study. The study involves running the scaling law analysis on models trained with different combinations of our synthetic data curriculum: (1) UIH data only, (2) CF and UIH data, and (3) the complete mixture of Item-Text, CF, and UIH data.

Our hypothesis is that training on the complete, mixed dataset will yield the most efficient scaling, evidenced by a larger scaling exponent ($\alpha$) compared to the ablated conditions. Such a result would demonstrate a super-additive, or synergistic, effect, proving that the integrated knowledge from all layers enables the model to learn more efficiently than the sum of its parts.

\begin{figure}[H]
    \centering
  \begin{subfigure}{0.49\textwidth}
    \includegraphics[trim={859 641 0 0}, clip, width=1\textwidth]{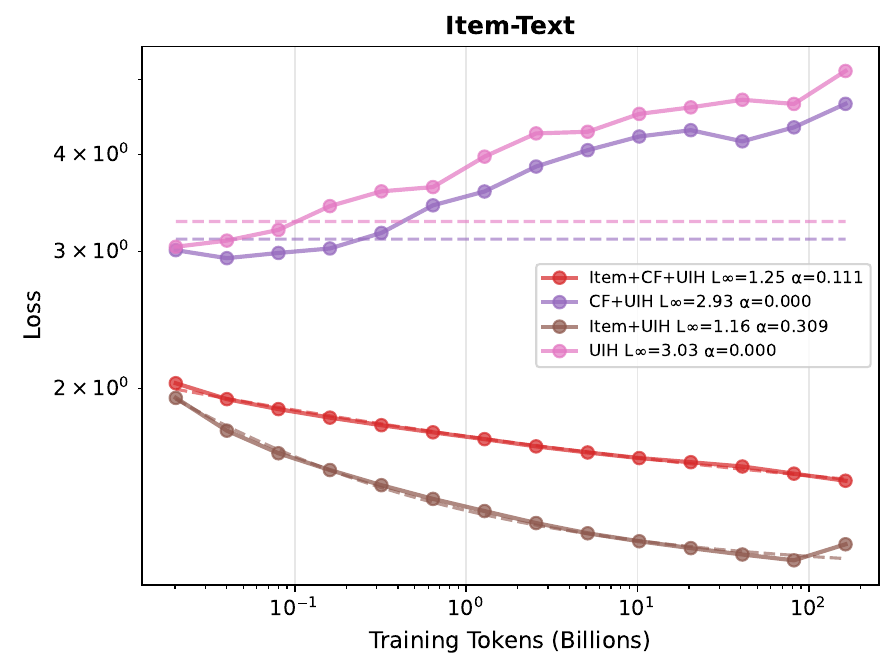}
  \end{subfigure}%
  \hfill
  \begin{subfigure}{0.49\textwidth}
    \includegraphics[trim={0 320 859 320}, clip, width=1\textwidth]{figure/domain_ablation.pdf}
  \end{subfigure}%
  \vfill
  \begin{subfigure}{0.49\textwidth}
    \includegraphics[trim={429 320 429 320}, clip, width=1\textwidth]{figure/domain_ablation.pdf}
  \end{subfigure}%
  \hfill
  \begin{subfigure}{0.49\textwidth}
    \includegraphics[trim={859 320 0 320}, clip, width=1\textwidth]{figure/domain_ablation.pdf}
  \end{subfigure}%
    \caption{Ablation studies on selection of domains for recommendation data. Obviously, excluding a domain will causes degradation on the corresponding domain.}
    \label{fig:domain_ablation}
\end{figure}

 To understand the contribution of each data domain, we conduct ablation studies by training models with different combinations of item-text, collaborative filtering (CF), and user interaction history (UIH) data. We evaluate perplexity across multiple test sets and fit scaling laws to characterize asymptotic performance. Table~\ref{tab:domain_ablation_data} provides mixture ratio of each domain, the other settings are exactly same. Figure~\ref{fig:domain_ablation} shows perplexity on hold-out evaluation dataset for each domain.

\begin{figure}[H]
    \centering
    \includegraphics[trim={0 0 429 641}, clip, width=0.9\textwidth]{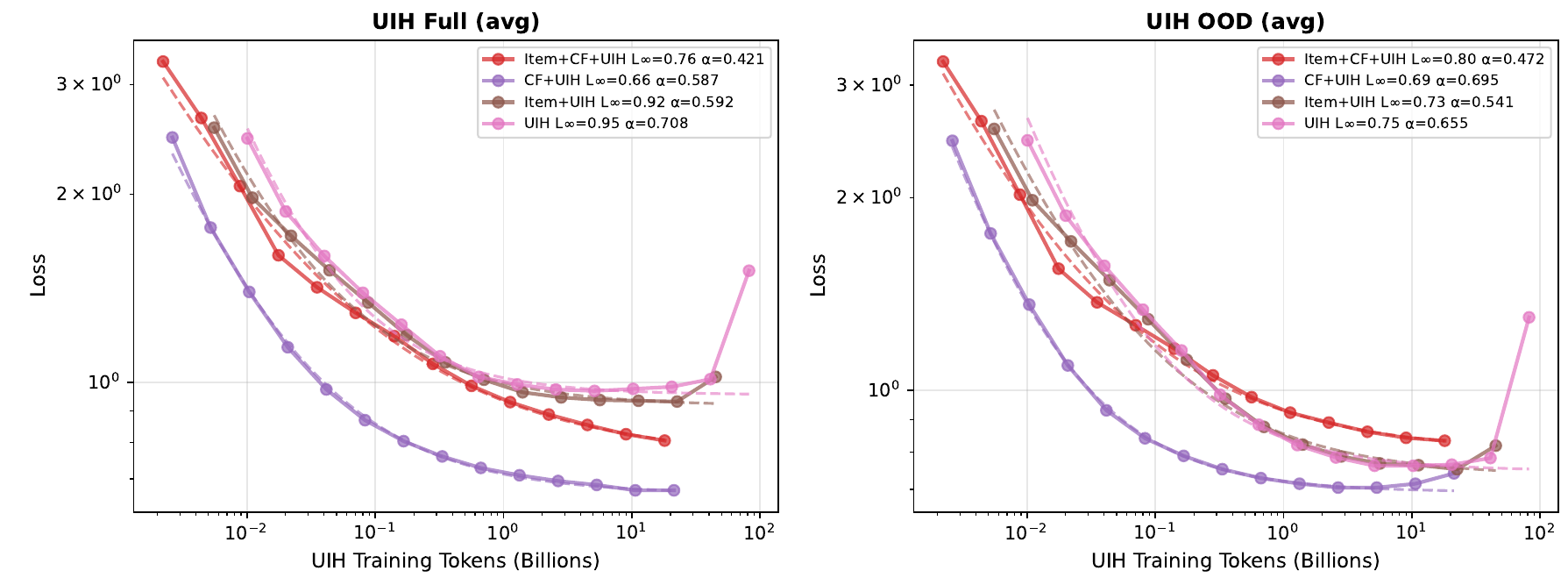}
    \caption{We plot the evaluation perplexity for UIH using number of UIH training tokens as X-axis. This figure indicates including CF data (Item-text + CF + UIH and CF + UIH) enables the model to learn UIH better.}
    \label{fig:uih_domain_ablation}
\end{figure}

 \textbf{Asymmetric Transfer Between CF and UIH}. We observe a notable asymmetry in cross-domain transfer. Including CF data alongside UIH significantly improves UIH modeling performance (Figure~\ref{fig:uih_domain_ablation}, with CF+UIH achieving the lowest asymptotic perplexity ($L_\infty=0.66$) on UIH evaluation sets—even outperforming the UIH-only baseline ($L_\infty=0.95$). This suggests that collaborative filtering signals, which capture user-item affinity patterns, provide complementary information that benefits sequential user behavior modeling. In contrast, the reverse transfer does not hold: training on UIH data alone yields virtually no improvement on CF tasks ($\alpha\approx 0$), indicating that user interaction sequences do not implicitly encode the pairwise preference signals required for collaborative filtering.

 \textbf{Domain-Specific Data Remains Essential}. Despite the positive transfer from CF to UIH, we find that domain-specific training data remains critical for each task. Models trained without CF data plateau at substantially higher CF loss ($L_\infty=1.30$) compared to those with CF data ($L_\infty=0.35$). Similarly, excluding item-text data leads to severe degradation on item-text evaluation ($L_\infty\approx3$ vs. $L_\infty\approx1.2$), while models retaining item-text data maintain stable performance throughout training.

 \textbf{Trade-offs in Multi-Domain Training}. Including item-text data introduces a modest trade-off: while it prevents catastrophic forgetting of item semantics, it slightly increases UIH loss ($L_\infty=0.76$ for Item+CF+UIH vs. $L_\infty=0.66$ for CF+UIH). This suggests that practitioners should consider their downstream priorities when designing data mixtures. More studies are provided in next section.

\begin{table}[]
\caption{Data mixture ratios of each domain for the experiments. For all experiments we keep 50\% for general text and all experiments are trained for 163B tokens.}
\label{tab:domain_ablation_data}
\centering
\begin{tabular}{l|lll}
\toprule
Choice\&Ratio & Item Text & CF     & UIH  \\
\midrule
\rowcolor{lightgray}
Item-text + CF + UIH  & 9\%    & 30\%   & 11\% \\
CF + UIH  & 0\%    & 37\%   & 13\% \\
Item-text UIH  & 22.5\%    & 0\%   & 27.5\% \\
UIH  & 0\%    & 0\%   & 50\% \\
\bottomrule
\end{tabular}
\end{table}

\subsection{Ablation Study: Data Mixtures}
\begin{figure}[H]
    \centering
  \begin{subfigure}{0.49\textwidth}
    \includegraphics[trim={859 680 0 0}, clip, width=1\textwidth]{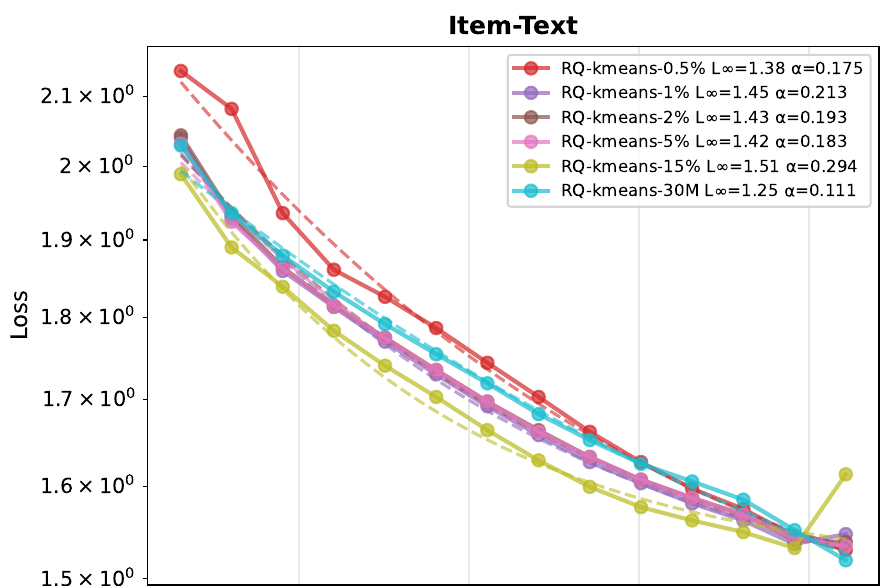}
  \end{subfigure}%
  \hfill
  \begin{subfigure}{0.49\textwidth}
    \includegraphics[trim={859 360 0 320}, clip, width=1\textwidth]{figure/uih_mixture_comparison.pdf}
  \end{subfigure}%
  \vfill
  \begin{subfigure}{0.49\textwidth}
    \includegraphics[trim={0 0 849 641}, clip, width=1\textwidth]{figure/uih_mixture_comparison.pdf}
  \end{subfigure}%
  \hfill
  \begin{subfigure}{0.49\textwidth}
    \includegraphics[trim={429 0 429 641}, clip, width=1\textwidth]{figure/uih_mixture_comparison.pdf}
  \end{subfigure}%
    \caption{Scaling laws on 4B models with different mixture ratio on UIH. Obviously the perplexity starts to increase when the UIH mixture ratio is too high (reduced UIH data is repeated too many times), which is a sign of overfitting. The higher mixture ratio, this increase happens earlier in training stage. We omitted the figures for CF Both Seen and CF On Unseen here, as they are very similar to CF Both Unseen.}
    \label{fig:repetition_analysis}
\end{figure}

A balanced mixture of different layers of data is also important to scaling of LLM training. To this end, we sample 42M tokens from 2.7B (or 39K samples from 5.748M samples) of our UIH data as a reduced UIH data. we then tested the UIH mixture ratio in 0.5\%, 1\%, 2\%, 5\% and 15\% for this reduced UIH data, the mixture ratio of other recommendation domain is normalized accordingly to maintain a sum of 50\%. As the UIH mixture ration on this reduced set increases, the more repeats of the reduced UIH data the model would go through (Column \#repeats of Table~\ref{tab:repetition_analysis}). In this study, we focus on 4B model and results are shown in Figure~\ref{fig:repetition_analysis} and Table~\ref{tab:repetition_analysis}.

\begin{figure}[H]
    \centering
  \begin{subfigure}{0.45\textwidth}
    \includegraphics[width=1\textwidth]{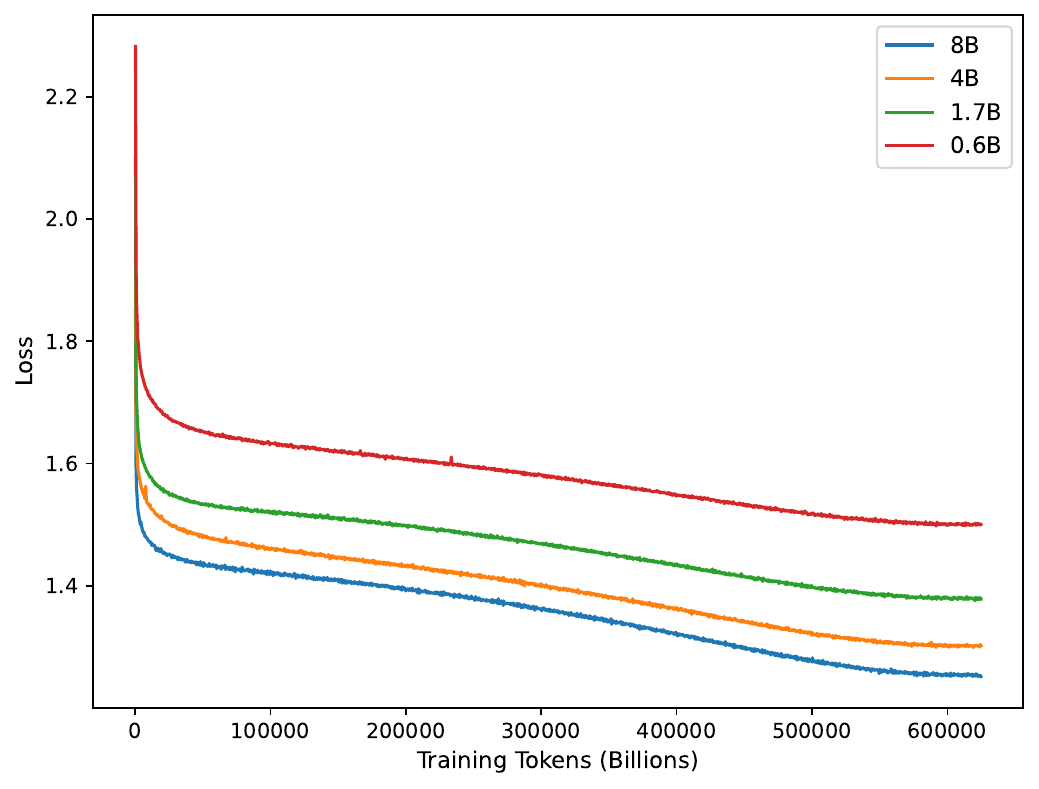}
  \end{subfigure}%
  \hfill
  \begin{subfigure}{0.45\textwidth}
    \includegraphics[width=1\textwidth]{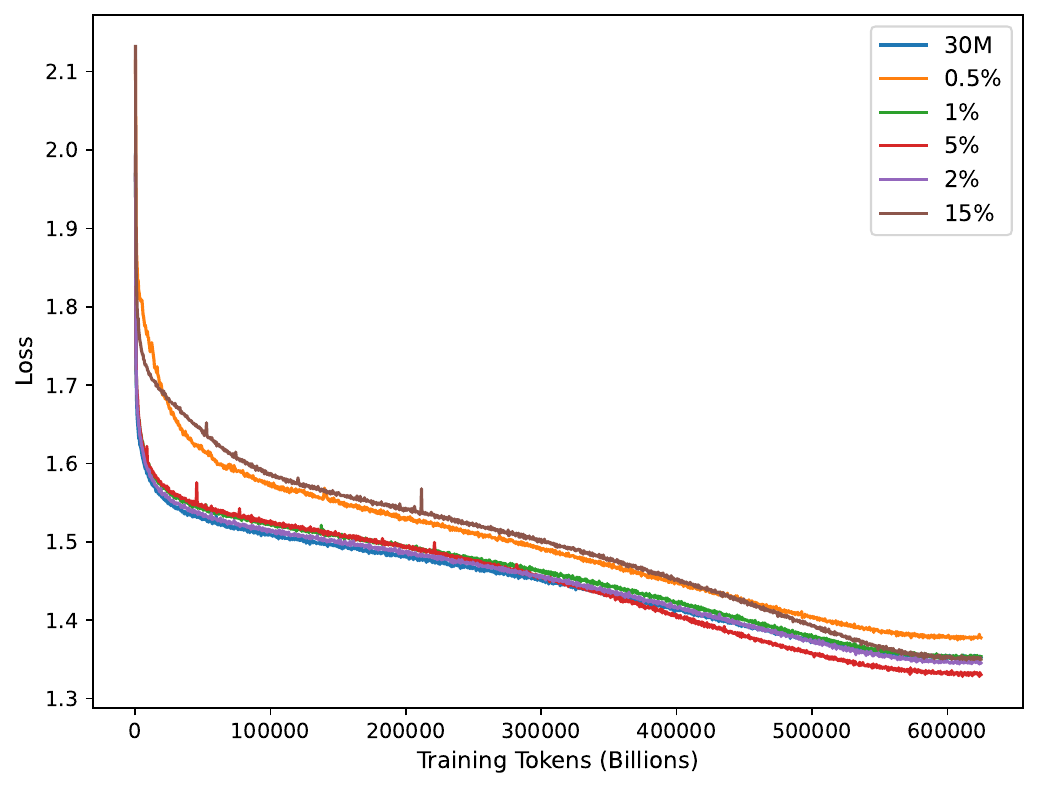}
  \end{subfigure}%
    \caption{(a) The training loss curve for experiment in Figure~\ref{fig:model_scale}. All the models are trained smoothly and show a monotonically decreasing training loss. (b) the training loss curve for experiment in Figure~\ref{fig:repetition_analysis}. The final training loss decreases in the order of $0.5\% > 1\% \approx 30M \approx 15\% > 2\% > 5\%$.}
    \label{fig:uih_loss}
\end{figure}

\begin{figure}[H]
  \centering
  \begin{subfigure}{1\textwidth}
    \includegraphics[trim={0 40 429 641}, clip, width=0.9\textwidth]{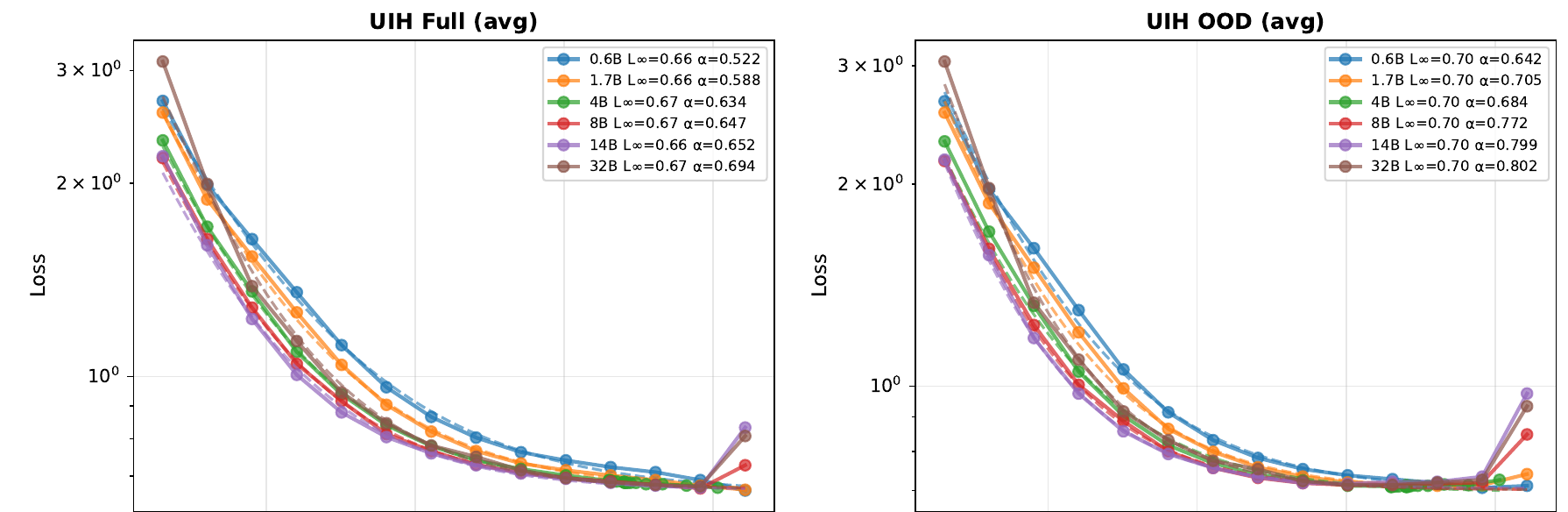}
    \caption{UIH ratio=2\%}
  \end{subfigure}%
  \vfill
  \begin{subfigure}{1\textwidth}
    \includegraphics[trim={0 0 429 641}, clip, width=0.9\textwidth]{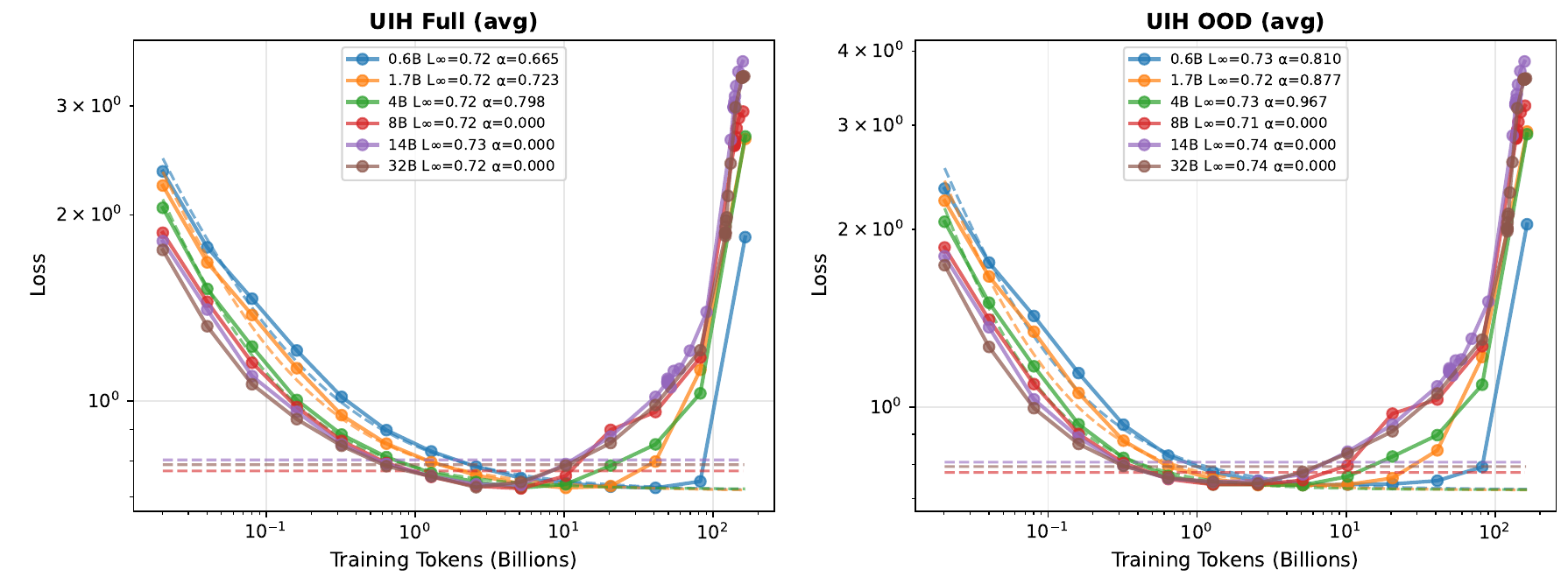}
    \caption{UIH ratio=15\%}
  \end{subfigure}%
    \caption{Scaling laws with mixture ratio on the reduced UIH data = (a) 2\% and (b) 15\% across different model scales. When the perplexity starts to increase depends on UIH mixture ratio, independent of model scales. We omitted the results on other domains as they shown the same patterns as Figure~\ref{fig:repetition_analysis}.}
    \label{fig:repetition_15p}
\end{figure}

\begin{table}[h]
\caption{Scaling laws on 4B models with different mixture ratio on UIH. We also provide the number of training tokens and repeats on the UIH data for a better understanding here. We provide mixture ratio for each domain in the left five columns and scaling law in the remaining 8 columns. The first row shows the result using full UIH dataset, which is set up in Figure~\ref{fig:model_scale}; the other rows are for reduced UIH dataset. $L_\infty$ should be the lower the better and $\alpha$ should be the higher the better.}
\label{tab:repetition_analysis}
\centering
\scriptsize
\begin{tabular}{lllll|llllllll}
\toprule
\multicolumn{3}{c|}{UIH}      & Item Text & CF & \multicolumn{2}{c}{General}                      & \multicolumn{2}{c}{Item-Text}   & \multicolumn{2}{c}{CF}     & \multicolumn{2}{c}{UIH}    \\
\%    & \#tokens & \#repeats & \%        & \% & $L_\infty\ \downarrow$ & $\alpha\ \uparrow$ & $L_\infty\ \downarrow$ & $\alpha\ \uparrow$ & $L_\infty\ \downarrow$ & $\alpha\ \uparrow$ & $L_\infty\ \downarrow$ & $\alpha\ \uparrow$ \\
\midrule
\rowcolor{lightgray}
11.0\% & 17.93B & 6.1   & 9.0\% & 30.0\% & 0.83 & 0.024 & 1.25 & 0.111 & 0.35 & 0.332 & 0.76 & 0.421 \\
0.5\%  & 0.82B  & 4.1   & 13.5\% & 36.0\% & 0.84 & 0.034 & 1.38 & 0.175 & 0.35 & 0.317 & 0.75 & 0.423 \\
1.0\%  & 1.63B  & 8.2   & 13\% & 36.0\% & 0.83 & 0.024 & 1.45 & 0.213 & 0.37 & 0.391 & 0.78 & 0.485 \\
2.0\%  & 3.26B  & 16.3  & 12.0\% & 36.0\% & 0.83 & 0.024 & 1.43 & 0.193 & 0.37 & 0.394 & 0.81 & 0.576 \\
5.0\%  & 8.15B  & 40.8  & 12.0\% & 33.0\% & 0.83 & 0.023 & 1.42 & 0.183 & 0.37 & 0.383 & 0.84 & 0.685 \\
15.0\% & 24.45B & 122.3 & 20.0\% & 15.0\% & 0.85 & 0.018 & 1.51 & 0.294 & 0.35 & 0.287 & 0.90 & 0.754\\
\bottomrule
\end{tabular}
\end{table}

\textbf{Analysis}: Figure~\ref{fig:repetition_analysis} indicates starting from UIH mixture ratio=2\% (about 16 repeats at the end of training according to Table\ref{tab:repetition_analysis}), the model's perplexity on the hold-out evaluation set starts to increase, which suggests an overfit is happening (according to Figure~\ref{fig:uih_loss}(b), the training loss decreases monotonically). Increasing ratio would cause the overfit happens even earlier: at 15\%, it happens at 20B training tokens (about 16 repeats); and at 5\%, it happens at 80B training tokens (about 20 repeats).

This result generally aligns with \cite{yang2024syntheticcontinuedpretraining} that given a well constructed synthetic dataset, CPT with 4 repeats could enable the model to learn the knowledge, and \cite{muennighoffScalingDataConstrainedLanguage2025} training beyond four repeats gives diminished returns, continuing for 20+ repeats could \cite{hronTrainingLanguageModels2024} decrease ability to generalize unseen data. \cite{allenzhu2024physicslanguagemodels33, yang2024syntheticcontinuedpretraining}, Performance degrades sharply when repetitions exceed approximately 100$\times$ the original dataset size.

We further analyze model scales affect this behavior with two UIH mixture ratio=(a) 2\% and (b) 15\% on the reduced UIH data. Figure~\ref{fig:repetition_15p} indicates when the perplexity starts to increase depends on UIH mixture ratio and independent of model scales: the perplexity on evaluation set starts to increase at 20B total training tokens (about 16 repeats on reduced UIH data) when ratio=15\%; or at 160B tokens (also about 16 repeats) when ratio=2\%, regardless of model scales. However it becomes more significant as the model gets larger.

\subsection{Scaling Laws Across Model and Data Scale}
The initial results establish that predictable scaling is achievable. The next critical step is to characterize these scaling laws across the two primary dimensions of LLM development—data and model size—to identify compute-optimal training regimes. This endeavor seeks to replicate the rigorous, multi-faceted scaling analysis of the Chinchilla paper within the recommendation domain, an objective made possible by our scalable source of high-quality data.

\begin{figure}[H]
    \centering
    \includegraphics[width=0.9\textwidth]{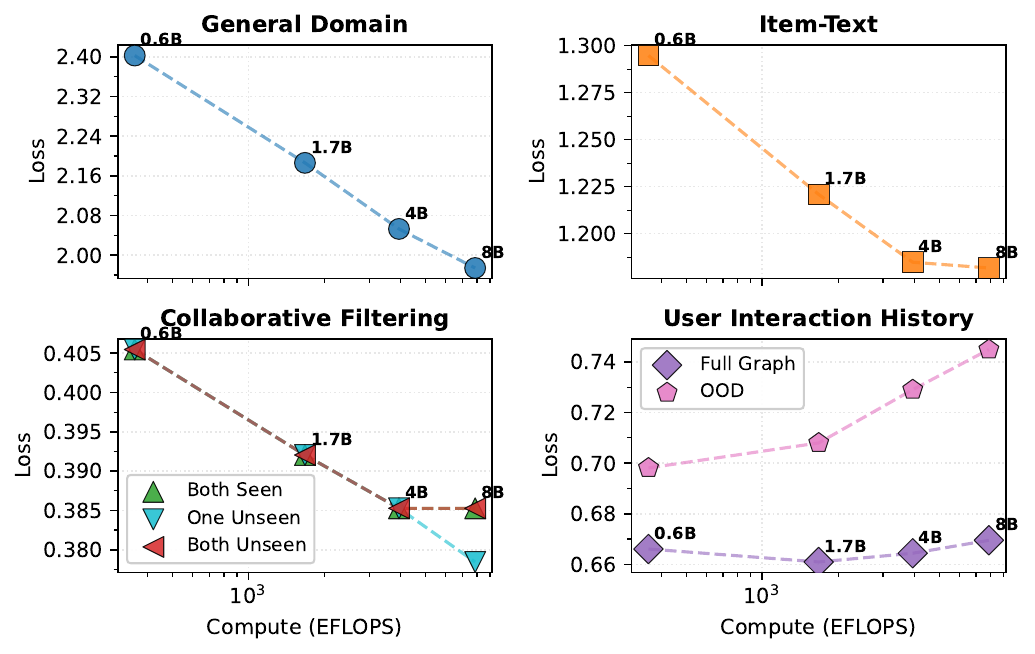}
    \caption{Compute-optimal analysis for models from 0.6B to 8B. Our scaling experiments train models from 0.6B to 8B parameters on 163.84B tokens, spanning from heavily over-trained (454 tokens/param for 0.6B) to near Chinchilla-optimal (20 tokens/param for 8B). While general-domain perplexity follows expected compute-optimal scaling, recommendation tasks—where domain-specific data is repeated 6–7× due to limited unique tokens—exhibit divergent behavior.
    \jh{In the graph, can you overlay the dots over the red line, and keep the same sig figs for all the y-axes? otherwise it is hard to see...}}
    \label{fig:compute_optimal}
\end{figure}

We conducted compute-optimal scaling analysis across four model sizes (0.6B–8B parameters), all trained on 163.84B tokens with a mixed-domain curriculum (shown in Figure~\ref{fig:compute_optimal}). This configuration places models at varying positions relative to Chinchilla-optimal allocation: the 8B model operates near the optimal 20 tokens-per-parameter ratio, while smaller models are progressively over-trained (0.6B at 454× tokens/param). For general-domain evaluation, where training data is not repeated, we observe classic Chinchilla scaling with a clear Pareto frontier—the 8B model achieves the lowest perplexity at highest compute.

However, recommendation-specific domains exhibit markedly different behavior. Collaborative filtering tasks saturate at model with 4B parameters (~1.47 PPL), indicating that the pattern complexity is bounded and fully learnable by 4B-parameter models given sufficient repetition. Most notably, user interaction history evaluation on out-of-distribution items reveals inverse scaling: the heavily over-trained 0.6B model (~2.01 PPL) outperforms the near-optimal 8B model (2.10 PPL). We attribute this to the interaction between data repetition and model capacity—larger models more readily memorize repeated behavioral sequences, leading to poorer OOD generalization, while the extreme over-training regime of smaller models appears to act as implicit regularization against such memorization.

\section{Case Studies}
In this section, we provide some examples of inferring the trained models. Figure~\ref{case:recommendation} shows how the model recommend an item given the UIH.
\begin{figure}[H]
  \begin{tcolorbox}[title={\textbf{Case Study: Recommendation given UIH}},
                    colback=white, colframe=black!70]
  \begin{minipage}[t]{0.48\textwidth}
  \textbf{\underline{Semantic ID Format}}\\[0.5em]
  \textbf{Prompt:}\\
  {\footnotesize\texttt{A user interacted with the following sequence of items: <RECTOKEN> REC2981 REC2337 REC5067 REC1796 REC2985 REC6054 REC988 REC5454 </RECTOKEN>, <RECTOKEN> REC2284 REC2581 REC2334 REC8032 REC6083 REC4502 REC5357 REC7604 </RECTOKEN>, <RECTOKEN> REC5411 REC2334 REC6499 REC5622 REC2394 REC6054 REC7338 REC7704 </RECTOKEN>, <RECTOKEN> REC6594 REC2334 REC5261 REC4570 REC6576 REC906 REC6676 REC6499 </RECTOKEN>, <RECTOKEN> REC2284 REC641 REC8166 REC7604 REC2581 REC2312 REC7393 REC1231 </RECTOKEN>, <RECTOKEN> REC5824 REC2334 REC4570 REC919 REC6513 REC2773 REC3722 REC7587 </RECTOKEN>, <RECTOKEN> REC5824 REC7479 REC2334 REC5622 REC6499 REC344 REC3722 REC1553 </RECTOKEN>}}\\[0.3em]
  \textbf{Response:} {\footnotesize\texttt{REC5824 REC7479 REC2334 REC5622 REC2006 REC6499 REC1553 REC344}}\\
  \textbf{Ground Truth:} {\footnotesize\texttt{REC5824 REC7479 REC2334 REC344 REC5622 REC2006 REC6499 REC3722}}
  \end{minipage}
  \hfill
  \begin{minipage}[t]{0.48\textwidth}
  \textbf{\underline{Decoded Natural Language}}\\[0.5em]
  \textbf{Prompt:}\\
  {\footnotesize A user interacted with the following sequence of items: [redacted] top, by [redacted] in Women › Athletic apparel › Athletic Tank Tops. Size: XS (0-2)., [redacted] Dress, by [redacted] in Women › Dresses › Above knee, mini. Size: XS (0-2)., Bracelet, in Women › Jewelry › Bracelets., Necklace, in Women › Jewelry › Necklaces., Dress, in Women › Dresses › Knee-length. Size: S (4-6)., crossbody bag, in Women › Women's handbags › Crossbody Bags, [redacted] purse, by [redacted] in Women › Women's handbags › Shoulder Bags. Color: black.}\\[0.3em]
  \textbf{Response:} {\footnotesize [redacted] handbag, by [redacted] in Women › Women's handbags › Shoulder Bags.}\\
  \textbf{Ground Truth:} {\footnotesize [redacted] purse, by [redacted] in Women › Women's handbags › Shoulder Bags. Color: black/gold}
  \end{minipage}
  \end{tcolorbox}
  \caption{An example of using the model to recommend item given the UIH. On the left, we provide the input and output with semantic ID format; and decoded natural language on the right. In this example, the model's response match perfectly with ground truth, except missing on the color.}\label{case:recommendation}
  \end{figure}

Applying beam search in inference, the model could generate relevant and diverse responses, which is shown in Figure~\ref{case:beam_search}. \jh{I think you need to have a section in the appendix that briefly discusses inference-time hyperparameters (e.g., sampling temperature, top-k/top-p, maximum sequence limit, etc.)}
\begin{figure}[H]
  \begin{tcolorbox}[title={\textbf{Case Study: Beam Search}},
                    colback=white, colframe=black!70]
  \begin{minipage}[t]{0.48\textwidth}
  \textbf{\underline{Semantic ID Format}}\\[0.5em]
  \textbf{Prompt:}\\
  {\footnotesize\texttt{A user interacted with the following sequence of items: <RECTOKEN> REC3972 REC7479 REC906 REC641 REC6621 REC4502 REC5992 REC2888 </RECTOKEN>, <RECTOKEN> REC805 REC7607 REC2334 REC6499 REC1209 REC2633 REC6341 REC3072 </RECTOKEN>, <RECTOKEN> REC3972 REC2590 REC2334 REC641 REC5454 REC5897 REC7479 REC4570 </RECTOKEN>, <RECTOKEN> REC2284 REC7604 REC641 REC2898 REC6581 REC2312 REC2581 REC748 </RECTOKEN>, <RECTOKEN> REC2284 REC641 REC7604 REC6581 REC2898 REC2581 REC8166 REC748 </RECTOKEN>, <RECTOKEN> REC2284 REC641 REC2581 REC8166 REC748 REC2312 REC1231 REC7604 </RECTOKEN>, <RECTOKEN> REC3972 REC6581 REC7479 REC2410 REC641 REC8174 REC3722 REC1231 </RECTOKEN>, <RECTOKEN> REC3375 REC2913 REC5824 REC7366 REC2752 REC984 REC1796 REC7604 </RECTOKEN>, <RECTOKEN> REC5824 REC82 REC5024 REC7427 REC1030 REC906 REC3968 REC2269 </RECTOKEN>, <RECTOKEN> REC5824 REC2334 REC7479 REC1776 REC1553 REC2269 REC5622 REC6499 </RECTOKEN>}}\\[0.3em]
  \textbf{Response 1:} {\footnotesize\texttt{REC5824 REC7479 REC2334 REC5024 REC82 REC6513 REC7736 REC1030}}\\
  \textbf{Response 2:} {\footnotesize\texttt{REC5824 REC7479 REC2334 REC5011 REC5622 REC2006 REC1553 REC3722}}\\
  \textbf{Response 3:} {\footnotesize\texttt{REC5824 REC7479 REC2334 REC5622 REC6499 REC344 REC3722 REC1553}}\\
  \end{minipage}
  \hfill
  \begin{minipage}[t]{0.48\textwidth}
  \textbf{\underline{Decoded Natural Language}}\\[0.5em]
  \textbf{Prompt:}\\
  {\footnotesize A user interacted with the following sequence of items: [redacted] Tote, by [redacted] in Women › Women's handbags › Tote Bags, Ring, in Women › Jewelry › Rings, [redacted], by [redacted] in Beauty › Skin care › Body, [redacted] Brazilian Crush, by [redacted] in Beauty › Fragrance › Women, [redacted] daisy, by [redacted] in Beauty › Fragrance › Women, crossbody bag, in Women › Women's handbags › Crossbody Bags, [redacted] Purse, by [redacted] in Women › Women's handbags › Crossbody Bags. Color: other, [redacted] crossbody Medium purse, by [redacted] in Women › Women's handbags › Crossbody Bags, [redacted] cross body bag, by [redacted] in Women › Women's handbags › Crossbody Bags. Color: other, [redacted] purse, by [redacted] in Women › Women's handbags › Satchel.}\\[0.3em]
  \textbf{Response 1:} {\footnotesize\texttt{[redacted] Field tote, by [redacted] in Women › Women's handbags › Shoulder Bags.}}\\
  \textbf{Response 2:} {\footnotesize\texttt{[redacted] Purse, by [redacted] in Women › Women's handbags › Shoulder Bags. Color: brown.}}\\
  \textbf{Response 3:} {\footnotesize\texttt{[redacted] purse, by [redacted] in Women › Women's handbags › Shoulder Bags. Color: black.}}\\
  \end{minipage}
  \end{tcolorbox}
  \caption{An example of applying beam search on the model's inference. On the left, we provide the input and output with semantic ID format; and decoded natural language on the right. In this example, the model's responses are relevant and diverse.}\label{case:beam_search}
  \end{figure}

It would be interesting to find out how the model would react to a random UIH. We provide such example in Figure~\ref{case:random_uih}, which shows when given a random UIH, the model still tries to recommendation items relevant to items to UIH; but given the items in the UIH are random, the model's recommendations are more diverse.
\begin{figure}[H]
  \begin{tcolorbox}[title={\textbf{Case Study: Recommendation given a Random UIH}},
                    colback=white, colframe=black!70]
  \begin{minipage}[t]{0.48\textwidth}
  \textbf{\underline{Semantic ID Format}}\\[0.5em]
  \textbf{Prompt:}\\
  {\footnotesize\texttt{A user interacted with the following sequence of items: <RECTOKEN> REC3311 REC953 REC418 REC7643 REC5624 REC419 REC7078 REC6875 </RECTOKEN>, <RECTOKEN> REC6812 REC6264 REC3311 REC7338 REC844 REC1746 REC237 REC5624 </RECTOKEN>, <RECTOKEN> REC6875 REC3311 REC2145 REC6576 REC5552 REC1796 REC2136 REC4685 </RECTOKEN>, <RECTOKEN> REC729 REC6261 REC4334 REC2044 REC1684 REC3927 REC4029 REC507 </RECTOKEN>, <RECTOKEN> REC5411 REC985 REC2394 REC7373 REC7338 REC4863 REC7366 REC6264 </RECTOKEN>, <RECTOKEN> REC7795 REC3245 REC856 REC6326 REC3812 REC6566 REC6855 REC7479 </RECTOKEN>, <RECTOKEN> REC3972 REC2145 REC7479 REC5454 REC5885 REC4540 REC6914 REC3812 </RECTOKEN>, <RECTOKEN> REC1166 REC237 REC984 REC1796 REC1295 REC5947 REC6341 REC6676 </RECTOKEN>, <RECTOKEN> REC6581 REC641 REC95 REC4647 REC5959 REC4208 REC4531 REC3722 </RECTOKEN>, <RECTOKEN> REC5695 REC4190 REC6261 REC2269 REC5624 REC1127 REC1796 REC729 </RECTOKEN>}}\\[0.3em]
  \textbf{Response 1} {\footnotesize\texttt{REC4905 REC2334 REC1098 REC3339 REC6326 REC4190 REC2269 REC4863}}\\
  \textbf{Response 2} {\footnotesize\texttt{REC641 REC2981 REC7479 REC524 REC6631 REC2334 REC2581 REC8032}}\\
  \textbf{Response 3} {\footnotesize\texttt{REC641 REC2981 REC5980 REC4071 REC95 REC4531 REC2479 REC1685}}\\
  \textbf{Response 4} {\footnotesize\texttt{REC641 REC2981 REC7479 REC524 REC6631 REC2334 REC573 REC3812}}\\
  \textbf{Response 5} {\footnotesize\texttt{REC641 REC2981 REC7479 REC524 REC6631 REC2334 REC573 REC409}}\\
  \end{minipage}
  \hfill
  \begin{minipage}[t]{0.48\textwidth}
  \textbf{\underline{Decoded Natural Language}}\\[0.5em]
  \textbf{Prompt:}\\
  {\footnotesize A user interacted with the following sequence of items: Play Air Maintenance Station Toy Plane, in Toys \& Collectibles › Sports \& Outdoor Play › Sand \& Water Toys, [redacted] Star Wars Snowspeeder Pilot Light Gray Helmet, by [redacted] in Toys \& Collectibles › Building Toys › [redacted] Toys, [redacted] Wooden Railway Bill and Ben tank engine wood train 23102701, by [redacted] \& Friends in Toys \& Collectibles › Remote Control Toys \& Vehicles › Trains \& Train Sets, [redacted] Flight Suit Mens Medium Large Costume Bradley Bradshaw Costume Halloween, by [redacted] in Men › Coats \& jackets › Military. Size: M (38-40), Rainbow Eraser Multicolor Charm Handmade Friendship Bracelet, by Handmade in Women › Jewelry › Bracelets, <Fantastic four [redacted] visionaries John Byrne, by [redacted] in Books › Fiction Books › Comics, Flia Shoes, in Kids › Girls shoes, samsung galaxy s8 [redacted] case. used, by [redacted] in Electronics › Cell phones \& accessories › Cases, covers \& skins, Floral Cardigan by [redacted], by [redacted] in Women › Sweaters › Cardigan. Size: M (8-10), VTG Sno Rider Womens Snow Pants Black 12 Insulated Bibs Snowmobile Zip up Legs, in Sports \& outdoors › Apparel › Women. Size: L (12-14).}\\[0.3em]
  \textbf{Response 1} {\footnotesize\texttt{Vintage 90s Ford Mustang Overalls, by Ford in Men › Pants › Other. Size}}\\
  \textbf{Response 2} {\footnotesize\texttt{Princess Polly, by Princess Polly in Women › Dresses › Above knee, mini. Size: M (}}\\
  \textbf{Response 3} {\footnotesize\texttt{Women’s size large tops bundle, by Boutique in Women › Tops \& blouses › Blouse.}}\\
  \textbf{Response 4} {\footnotesize\texttt{Dollskill x Alice in Wonderland, by Dolls Kill in Women › Tops \& blouses › Other}}\\
  \textbf{Response 5} {\footnotesize\texttt{Dolls Kill x The Powerpuff Girls - Bubbles \& Butterflies Corset, by Doll}}\\
  \end{minipage}
  \end{tcolorbox}
  \caption{An example of model makes recommendation given a random UIH. On the left, we provide the input and output with semantic ID format; and decoded natural language on the right. In this example, the model's responses are still relevant to the items presented in the UIH.}\label{case:random_uih}
  \end{figure}

Finally we provide an example of retrieving top-10 similar items given an generated item with semantic ID in Table~\ref{tab:example_retrieval}. This example shows the effectiveness of the semantic ID in representing the items and capture their semantics.
\begin{table}[]
\caption{Example of running similarity search based on semantic IDs. Given the query (Row 1), we retrieval top-10 most similar items.}
\label{tab:example_retrieval}
\centering
\begin{tabular}{p{0.8\textwidth}|l}
\toprule
Query  & Similarity    \\
\midrule
\rowcolor{lightgray}
REC5824 REC2334 REC4570 REC3722 REC7479 REC3968 REC2388 REC6054 \newline Handbag, in Women › Women's handbags › Shoulder Bags & Query \\
\midrule
REC5824 REC2334 REC4570 REC3722 REC7479 REC3968 REC2388 REC6054 \newline Handbag, in Women › Women's handbags › Shoulder Bags & 100.0\% \\\hline
REC5824 REC2334 REC4570 REC3722 REC7479 REC3968 REC2388 REC2240 \newline Beautiful Bag Designer Style!!, in Women › Women's handbags › Shoulder Bags & 87.5\%\\\hline
REC5824 REC2334 REC4570 REC3722 REC7479 REC3968 REC2388 REC7587 \newline Light Gold Tote, in Women › Women's handbags › Shoulder Bags & 87.5\% \\\hline
REC5824 REC2334 REC4570 REC3722 REC7479 REC3968 REC2388 REC984 \newline luxury bags, in Women › Women's handbags › Shoulder Bags & 87.5\% \\\hline
REC5824 REC2334 REC4570 REC3722 REC7479 REC3968 REC1030 REC2633 \newline inspired bag, in Women › Women's handbags › Shoulder Bags & 75.0\% \\\hline
REC5824 REC2334 REC4570 REC3722 REC7479 REC3968 REC1228 REC5454 \newline shoes women, in Women › Women's handbags › Crossbody Bags & 75.0\% \\\hline
REC5824 REC2334 REC4570 REC3722 REC7479 REC3968 REC1553 REC6842 \newline Stay Real Bag, in Women › Women's handbags › Shoulder Bags & 75.0\% \\\hline
REC5824 REC2334 REC4570 REC3722 REC7479 REC3968 REC2240 REC1228 \newline Fancy Handbag, in Women › Women's handbags › Shoulder Bags & 75.0\% \\\hline
REC5824 REC2334 REC4570 REC3722 REC7479 REC3968 REC2240 REC2388 \newline beautiful bag, by Op in Women › Women's handbags › Shoulder Bags &  75.0\% \\\hline
REC5824 REC2334 REC4570 REC3722 REC7479 REC3968 REC2240 REC4075 \newline Stylish Purse, Statement Piece, in Women › Women's handbags › Shoulder Bags & 75.0\% \\
\bottomrule
\end{tabular}
\end{table}

\section{Discussion and Conclusion}
This paper has addressed a foundational gap in the application of Large Language Models to recommender systems. The absence of predictable scaling laws has, until now, hindered the scientific and systematic development of these powerful new models. The central contribution of this work is the demonstration that this barrier is not inherent to the domain but is a consequence of relying on pathologically flawed raw user data.

By introducing a novel, layered framework for generating high-quality synthetic data, this research has shown that it is possible to create a curated curriculum that enables robust, predictable, power-law scaling. Scaling exponents reveal that complex behavioral patterns benefit most from additional data (UIH: $\alpha$=0.45--0.59 vs.\ Item-Text: $\alpha$=0.15), while asymmetric cross-domain transfer---where CF data accelerates UIH learning by 31\%---provides actionable guidance for optimal data mixing.

\subsection{Cross-Model and Cross-Dataset Generalization}
Since submission, we have extended our experiments to a large-scale proprietary social media recommendation dataset with both the \textbf{Gemma 3} family (1B, 4B, 12B) and \textbf{Qwen3} (0.6B--30B-A3B). Both families exhibit stable power-law scaling with RMSE $<$ 0.035. As shown in Table~\ref{tab:gemma3_scaling}, the hierarchy of learning efficiency ($\alpha$: UIH $>$ Alignment $>$ General) is preserved, and $\alpha$ decreases monotonically with model size across all domains. We also extended the Qwen3-4B token budget from 163B to 1,104B tokens on this dataset, confirming that scaling behavior remains predictable with no saturation.

\begin{table}[h]
\centering
\caption{Fitted scaling law parameters for Gemma 3 on a proprietary social media dataset. All fits have RMSE $<$ 0.035. The hierarchy of learning efficiency ($\alpha$: UIH $>$ Alignment $>$ General) is preserved across model families. $\alpha$ denotes the scaling exponent (higher is better); $L_\infty$ denotes the asymptotic loss (lower is better).}
\label{tab:gemma3_scaling}
\begin{tabular}{l|ll|ll|ll}
\toprule
\textbf{Model} & \multicolumn{2}{c|}{\textbf{General}} & \multicolumn{2}{c|}{\textbf{Alignment}} & \multicolumn{2}{c}{\textbf{UIH}} \\
 & $L_\infty$ & $\alpha$ & $L_\infty$ & $\alpha$ & $L_\infty$ & $\alpha$ \\
\midrule
Gemma-3-1B & 1.097 & 0.025 & 1.292 & 0.255 & 1.246 & 0.274 \\
Gemma-3-4B & 0.952 & 0.038 & 0.635 & 0.085 & 1.085 & 0.237 \\
Gemma-3-12B & 0.874 & 0.051 & 0.589 & 0.096 & 0.957 & 0.205 \\
\bottomrule
\end{tabular}
\end{table}

The fitted joint scaling laws for Qwen3 on the social media dataset are:
\begin{align}
  \ell_{\text{general}} &= 0.00 + 5.51\, N^{-0.262} + 1.51\, D^{-0.030} \\
  \ell_{\text{item-text}} &= 0.00 + 2.99\, N^{-0.169} + 0.73\, D^{-0.117} \\
  \ell_{\text{UIH}} &= 1.15 + 5.25\, N^{-0.380} + 6.72\, D^{-0.894}
\end{align}
UIH on the social media dataset exhibits even stronger data scaling ($\beta{=}0.894$) than on MerRec ($\beta{=}0.272$), while the inverse relationship between model-size and data scaling exponents is preserved, reinforcing the conclusion that recommendation domains are fundamentally data-dominant.

\subsection{Perplexity vs.\ Downstream Metrics}
While perplexity is our primary scaling metric, several lines of evidence support its practical relevance. The TSTR results (Figure~\ref{fig:tstr_vs_trtr_recall}) demonstrate that models trained on our synthetic data achieve superior ranking performance on real data, and our extended experiments on the proprietary social media dataset confirmed a strong correlation between perplexity and Recall@K / NDCG for next-item prediction. This is analogous to the Chinchilla scaling law methodology~\cite{hoffmann2022training}, which also uses loss-based scaling but became foundational for compute-optimal training. Production evaluation of the CPT-trained LLM on end-to-end recommendation metrics in a live system is underway.

\subsection{Signal Quality vs.\ Distributional Simplicity}
An important question is whether our improvements come from denoised signals or merely distributionally easier data. Several findings argue for the former: (i)~TSTR outperforms TRTR---if synthetic data were simply easier, models should overfit to its distribution and fail on real test data; (ii)~asymmetric cross-domain transfer (CF$\rightarrow$UIH but not UIH$\rightarrow$CF) is inconsistent with uniform simplicity; (iii)~domain-specific scaling exponents reflect information density, not uniform easiness---``easy'' data would show uniformly high $\alpha$; (iv)~excessive repetition degrades performance (Section~\ref{fig:repetition_analysis}), inconsistent with the ``easy distribution'' hypothesis, which would predict continued improvement with repetition.

\subsection{Implications for Practitioners and Researchers}
The findings presented in this paper carry significant practical and theoretical implications. For practitioners, the establishment of the first scaling laws for LLMs in recommendation provides an essential tool for project planning and resource management. It enables teams to transition from speculative, heuristic-based approaches to a more scientific method of forecasting the computational budget and data volume required to achieve a target level of performance, thereby making the development of large-scale systems more predictable and justifiable. Future work will extend this analysis to post-training stages (SFT, RLHF) and test-time compute, while validating downstream task metrics beyond perplexity.

\section*{Acknowledgments}
We thank the anonymous ICML reviewers for their constructive feedback, which significantly improved this paper.

  \section*{Impact Statement}

  This paper presents work whose goal is to advance the field of Machine Learning, specifically the development of LLMs for recommender systems. We highlight several aspects of broader impact.

  \textbf{Positive Societal Impacts.} Our synthetic data framework offers inherent privacy benefits: the generated user interaction histories are derived from aggregated item-to-item graphs rather than individual user traces, decoupling training data from real user behavior. Additionally, our methodology addresses key systemic biases (particularly position and presentation-order bias) that plague real-world recommendation systems. By training on position-debiased synthetic curricula with substantially more uniform item exposure (Gini 0.64 vs.\ $>$0.95 in real logs), LLMs may produce more equitable recommendations that surface long-tail and niche content.

  \textbf{Potential Risks and Mitigations.} More effective recommendation systems could, in principle, increase user engagement in ways that raise concerns about digital well-being. However, our work focuses on the \emph{predictability} of model development rather than optimizing for engagement metrics. Furthermore, while our synthetic data generation aims to remove systemic biases, practitioners must validate that downstream applications do not inadvertently introduce new biases through their choice of source data or graph construction. We encourage future work to develop evaluation frameworks that audit both the fidelity and fairness of synthetic recommendation data.


\clearpage
\newpage
\bibliographystyle{assets/plainnat}
\bibliography{paper}

\clearpage
\newpage
\beginappendix

\appendix
\section{Hyperparameter Tuning for Synthetic UIH Generation}\label{sec:uih_parameter}
To generate the highest quality Synthetic User Interaction Histories (UIH), we conducted a comprehensive hyperparameter sweep of our Node2Vec-based random walk algorithm. We evaluated 14 distinct configurations, including a 1st-order (DeepWalk-style) baseline and 12 variants of a 2nd-order, BFS-like strategy ($p=0.5, q=2.0$). The sweep varied two key parameters:
\begin{enumerate}
    \item $\alpha_{stop}$: The probability of terminating a walk at any given step.
    \item path\_conf\_threshold: A minimum cumulative path confidence threshold used to filter out low-quality sequences.
\end{enumerate}
\begin{table}[]
\caption{Experiments on different hyperparameters for Node2Vec-based random walk algorithm to construct synthetic UIH. Based on the metrics, we selected `1st\_order\_a015\_thresh1e-09`, which is highlighted in gray.}
\begin{tabular}{lllllllll}
\toprule
\textbf{Config}                 & \textbf{Alpha} & \textbf{Thresh} & \textbf{Length} & \textbf{Items ↑} & \textbf{Gini ↓} & \textbf{Tokens ↑} & \textbf{Token Gini ↓} & \textbf{Geom Lift ↑} \\
\midrule
 bfs\_p05q2\_a10\_threshNone   & 0.10           & None            & 14.0            & 189K             & 0.732           & 1,580             & 0.856                 & 353.00               \\
 bfs\_p05q2\_a15\_threshNone   & 0.15           & None            & 10.6            & 183K             & 0.697           & 1,577             & 0.851                 & 372.28               \\
 bfs\_p05q2\_a20\_threshNone   & 0.20           & None            & 9.0             & 179K             & 0.672           & 1,584             & 0.849                 & 385.22               \\
 bfs\_p05q2\_a10\_thresh1e-09  & 0.10           & 1e-09           & 8.6             & 178K             & 0.660           & 1,574             & 0.845                 & 368.91               \\
 bfs\_p05q2\_a15\_thresh1e-09  & 0.15           & 1e-09           & 8.1             & 176K             & 0.650           & 1,573             & 0.844                 & 384.71               \\
 bfs\_p05q2\_a20\_thresh1e-09  & 0.20           & 1e-09           & 7.6             & 174K             & 0.643           & 1,568             & 0.844                 & 388.84               \\
 bfs\_p05q2\_a10\_thresh2e-09  & 0.10           & 2e-09           & 8.4             & 177K             & 0.656           & 1,571             & 0.844                 & 375.29               \\
 bfs\_p05q2\_a15\_thresh2e-09  & 0.15           & 2e-09           & 8.0             & 175K             & 0.648           & 1,569             & 0.844                 & 384.51               \\
 bfs\_p05q2\_a20\_thresh2e-09  & 0.20           & 2e-09           & 7.5             & 174K             & 0.641           & 1,566             & 0.843                 & 390.54               \\
 bfs\_p05q2\_a10\_thresh3e-09  & 0.10           & 3e-09           & 8.3             & 177K             & 0.655           & 1,569             & 0.844                 & 376.11               \\
 bfs\_p05q2\_a15\_thresh3e-09  & 0.15           & 3e-09           & 7.9             & 175K             & 0.647           & 1,571             & 0.844                 & 384.48               \\
 bfs\_p05q2\_a20\_thresh3e-09  & 0.20           & 3e-09           & 7.5             & 174K             & 0.639           & 1,570             & 0.843                 & 392.59               \\
 1st\_order\_a015\_threshNone  & 0.15           & None            & 10.6            & 188K             & 0.680           & 1,579             & 0.850                 & 376.80               \\
 \rowcolor{lightgray}
 1st\_order\_a015\_thresh1e-09 & 0.15           & 1e-09           & 8.0             & 180K             & 0.635           & 1,581             & 0.844                 & 385.54      \\
 \bottomrule
\end{tabular}
\end{table}
\section{Methodology for Semantic Tokenization}\label{app:token}
This appendix details the methodology for determining the semantic item representation (<RECTOKEN>). As this representation serves as the fundamental vocabulary for our LLM, its quality is paramount. We evaluated three candidate strategies---a pre-trained Sparse Autoencoder (SAE), an in-domain Residual-Quantized VAE (RQ-VAE), and an in-domain RQ-kmeans---and selected SAE based on ablation experiments.
\subsection{Candidate Methodologies}
We evaluated two primary approaches for generating discrete item tokens:
\begin{itemize}
    \item \textbf{Approach 1: Pre-trained Sparse Autoencoder (SAE).}
This method utilizes a large-scale SAE originally trained on an internal multi-modal dataset comprising video, photo, and text data. The procedure involves passing the item's textual description through the pre-trained embedding model and then into the SAE. The top-k activated features (concepts) from the SAE are selected as the tokens.
\begin{itemize}
    \item \textbf{Potential Strength}: Leverage of a "foundation" encoder with exposure to massive, diverse multi-modal data.
    \item \textbf{Critical Risk (Semantic Mismatch)}: The SAE features were learned in a video-centric domain. Applying this vocabulary to e-commerce product text introduces a severe risk of negative transfer, where the "concepts" extracted are mathematically proximal but semantically incongruous for the recommendation task.
\end{itemize}
\item \textbf{Approach 2: In-Domain Residual-Quantized VAE (RQ-VAE).}
This method involves training a hierarchical RQ-VAE from scratch specifically on the target dataset (merrec). We utilize the Qwen2.5-7B model to generate high-quality text embeddings for all items. The RQ-VAE is then trained to quantize these embeddings into a sequence of discrete codes using a residual codebook structure.
\begin{itemize}
    \item \textbf{Potential Strength}: The learned vocabulary is perfectly aligned with the target domain's semantic distribution, as it is trained directly on the item descriptions.
    \item \textbf{Critical Risk (Data Leakage)}: A naive implementation that trains the tokenizer on the full dataset would leak test-set information into the training process, invalidating downstream scaling law experiments.
\end{itemize}
\item  \textbf{Approach 3: In-Domain RQ-kmeans.}
This method is similar to RQ-VAE but more stable and scalable. In our experiment we found using 6 layers of codebook and 256 codes per layer, the RQ-kmeans could reach collision rate 5.33\% but RQ-VAE is stuck at 50.54\%.
\end{itemize}

\subsection{Selected Methodology and Mitigation Protocol}
\begin{figure}[H]
    \centering
  \begin{subfigure}{0.48\textwidth}
    \includegraphics[trim={0 680 859 0}, clip, width=1\textwidth]{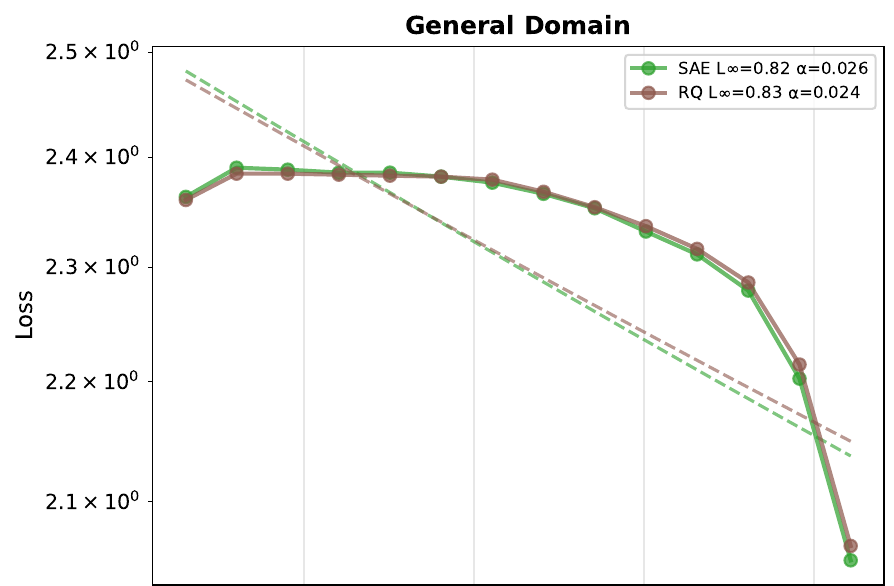}
  \end{subfigure}%
  \hfill
  \begin{subfigure}{0.48\textwidth}
    \includegraphics[trim={859 680 0 0}, clip, width=1\textwidth]{figure/tokenizer_comparison_4B.pdf}
  \end{subfigure}%
  \vfill
  \begin{subfigure}{0.37\textwidth}
    \includegraphics[trim={0 360 859 320}, clip, width=1\textwidth]{figure/tokenizer_comparison_4B.pdf}
  \end{subfigure}%
  \hfill
  \begin{subfigure}{0.31\textwidth}
    \includegraphics[trim={501 360 429 320}, clip, width=1\textwidth]{figure/tokenizer_comparison_4B.pdf}
  \end{subfigure}%
  \hfill
  \begin{subfigure}{0.31\textwidth}
    \includegraphics[trim={930 360 0 320}, clip, width=1\textwidth]{figure/tokenizer_comparison_4B.pdf}
  \end{subfigure}%
  \vfill
  \begin{subfigure}{0.54\textwidth}
    \includegraphics[trim={0 0 859 641}, clip, width=1\textwidth]{figure/tokenizer_comparison_4B.pdf}
  \end{subfigure}%
  \hfill
  \begin{subfigure}{0.45\textwidth}
    \includegraphics[trim={501 0 429 641}, clip, width=1\textwidth]{figure/tokenizer_comparison_4B.pdf}
  \end{subfigure}%
    \caption{Scaling laws on 4B models with two tokenization methods SAE vs RQ-kmeans. The other settings are exactly the same. SAE clearly outperformed RQ-kmeans across all domains.}
    \label{fig:socae_rq}
\end{figure}

We selected Approach 1 (SAE) as the our tokenization method after ablation against RQ-kmeans (6 layers of codebook and 256 codes per layer, after sweeping hyper-parameters Table~\ref{tab:rq_config}). Figure~\ref{fig:socae_rq} shows the comparison of scaling laws based on SAE vs RQ-Kmeans on 4B models. This figure shows SAE consistently outperformed RQ-Kmeans across all domains and steps.

\begin{table}[h]
\caption{RQ configuration sweep on Qwen-0.6B embeddings (64.1M items). All experiments use beam size 5 with progressive search and uniform sampling. Collision rate measured after uniform sampling.}
\label{tab:rq_config}
\centering
\small
\begin{tabular}{cccr}
\toprule
\textbf{Layers} & \textbf{Codes/Layer} & \textbf{Total Codes} & \textbf{Collision \%} \\
\midrule
3 & 256 & $256^3$ (16M) & 82.7 \\
3 & 512 & $512^3$ (134M) & 37.4 \\
5 & 256 & $256^5$ (1.1T) & 9.7 \\
6 & 256 & $256^6$ (281T) & \textbf{3.5} \\
8 & 128 & $128^8$ (72057T) & 1.2 \\
\bottomrule
\end{tabular}
\end{table}

\section{Quantitative Bias Audit}\label{app:bias_audit}
To rigorously characterize which biases our framework mitigates and which persist, we conducted a comprehensive analysis using the 14 hyperparameter configurations from Appendix~\ref{sec:uih_parameter} (100K sequences each). Table~\ref{tab:bias_audit} summarizes the distributional properties of synthetic UIH compared to real user logs.

\begin{table}[h]
\centering
\caption{Bias audit comparing synthetic UIH against real user logs. Item Gini coefficient measures item exposure inequality (lower = more uniform). Synthetic UIH achieves substantially more uniform item exposure than real logs.}
\label{tab:bias_audit}
\begin{tabular}{p{4cm}|p{5cm}|p{5cm}}
\toprule
\textbf{Metric} & \textbf{Synthetic UIH} & \textbf{Real Logs} \\
\midrule
Item Gini coefficient & 0.635--0.732 (prod config: 0.639) & $>$0.95 (top 1\% items get 80\% interactions) \\
\midrule
Unique items per 100K seqs & 174K--189K & --- \\
\midrule
Semantic token coverage & 88.6--89.5\% & --- \\
\midrule
Position bias & Zero by construction & CTR drops 50\%+ from pos.~1 to 5 \\
\midrule
Popularity bias & Partially mitigated (edge weights retain some popularity signal) & Strong feedback loops \\
\bottomrule
\end{tabular}
\end{table}

\subsection{Biases Mitigated vs.\ Retained}
Our framework eliminates two classes of bias by construction: (i)~\textbf{Positional bias}---the random walk has no concept of rank or presentation order; each step is determined solely by graph structure and $(p,q)$ parameters, ensuring zero positional artifacts. (ii)~\textbf{Temporal ordering bias}---the walk process generates sequences based on graph connectivity, not temporal ordering of real user sessions.

However, \textbf{popularity bias} encoded in the co-occurrence graph edge weights is only partially mitigated: popular items naturally have higher-weight edges and thus higher visitation probabilities during random walks. The moderate Gini coefficients (0.635--0.732 vs.\ near-1.0 in real logs) indicate that our framework substantially reduces but does not eliminate popularity concentration. This is an important distinction---we revise our framing from ``unbiased'' to ``position-debiased'' throughout the paper to accurately reflect these properties.

\subsection{Data Isolation Protocol}
We confirm full isolation across all experimental splits:
\begin{enumerate}
    \item The CF graph is constructed exclusively from training-split interactions. Test items are either held out entirely (CF Both Unseen) or partially (CF One Unseen), as defined in Appendix~\ref{sec:uih_parameter}.
    \item The SAE tokenizer is trained on training-split item descriptions only.
    \item UIH evaluation sequences contain no overlap with training UIH sequences.
    \item No test-derived graph structure is used anywhere outside evaluation.
\end{enumerate}

\end{document}